\documentclass[fleqn,usenatbib]{mnras}

\usepackage{newtxtext,newtxmath}

\usepackage[T1]{fontenc}

\DeclareRobustCommand{\VAN}[3]{#2}
\let\VANthebibliography\thebibliography
\def\thebibliography{\DeclareRobustCommand{\VAN}[3]{##3}\VANthebibliography}

\usepackage{graphicx}	
\usepackage{amsmath}	
\title[PIP shock stability]{Stability of two-fluid partially-ionised slow-mode shock fronts}

\author[B. Snow \& A. Hillier]{
B. Snow,$^{1}$\thanks{E-mail: b.snow@exeter.ac.uk}
and A. Hillier$^{1}$
\\
$^{1}$University of Exeter, UK
}

\date{Accepted XXX. Received YYY; in original form ZZZ}

\pubyear{2021}

\begin{document}
\label{firstpage}
\pagerange{\pageref{firstpage}--\pageref{lastpage}}
\maketitle

\begin{abstract}
   {A magnetohydrodynamic (MHD) shock front can be unstable to the corrugation instability, which causes a perturbed shock front to become increasingly corrugated with time. An ideal MHD parallel shock (where the velocity and magnetic fields are aligned) is unconditionally unstable to the corrugation instability, whereas the ideal hydrodynamic (HD) counterpart is unconditionally stable. For a partially ionised medium (for example the solar chromosphere), both hydrodynamic and magnetohydrodynamic species coexist and the stability of the system has not been studied.}
   {In this paper, we perform numerical simulations of the corrugation instability in two-fluid partially-ionised shock fronts to investigate the stability conditions, and compare the results to HD and MHD simulations. }
   {Our simulations consist of an initially steady 2D parallel shock encountering a localised upstream density perturbation. In MHD, this perturbation results in an unstable shock front and the corrugation grows with time. We find that for the two-fluid simulation, the neutral species can act to stabilise the shock front. A parameter study is performed to analyse the conditions under which the shock front is stable and unstable.}
   {We find that for very weakly coupled or very strongly coupled partially-ionised system the shock front is unstable, as the system tends towards MHD. However, for a finite coupling, we find that the neutrals can stabilise the shock front, and produce new features including shock channels in the neutral species. 
   We derive an equation that relates the stable wavelength range to the ion-neutral and neutral-ion coupling frequencies and the Mach number. Applying this relation to umbral flashes give an estimated range of stable wavelengths between 0.6 and 56 km. 
   }
\end{abstract}




\section{Introduction}

Magnetohydrodynamic (MHD) shock waves are a fundamental feature of astrophysical systems, occurring across a range of physical systems, for example, the solar atmosphere \citep{Beckers1969,Hollweg1982} and molecular clouds \citep{Draine1983}. 
The formation mechanism behind astrophysical shocks is as varied and results as a consequence of a number of physical processes, such as magnetic reconnection \citep{Yamada2010,Petschek1964} and wave steepening \citep{Suematsu1982}. 
For a hydrodynamic (HD) system, sonic shocks can exist as a transition from above to below the sound speed. In magnetohydrodynamics (MHD) there are three characteristic wave speeds (slow, Alfv\'en, fast) which leads to a wealth of possible shock transitions \citep[][]{Tidman1971}. Here we study the stability of slow mode shock fronts in partially ionised systems. 

For a steady-state shock, the MHD shock jumps can be described in terms of the conservative quantities sufficiently upstream and downstream of the shock. 
It can be shown that MHD systems support a variety of shock transitions (broadly categorised as slow, intermediate, and fast) as well as contact and tangential discontinuies.
If a non-zero component of magnetic field exists normal to the steady shock ($B_x \neq 0$) then 
jumps in the tangential magnetic field imply jumps in the tangential velocity. 
As such, vorticity ($\nabla \times \textbf{v}$) can exist at an MHD shock front. 
For an MHD contact discontinuity, where there is zero mass flux across the interface, only jumps in the density are permitted. As such, there is no magnetic field jump to support a jump in tangential velocity hence vorticity cannot exist at a steady-state MHD contact discontinuity.

In a hydrodynamic system, the magnetic field is zero and hence the shock jumps are easier to compute. 
With regards to vorticity, the HD system has the opposite behaviour to an MHD system; vorticity can be exist across a HD contact discontinuity, but not a HD shock \citep[since a transverse velocity is no longer supported by a magnetic field,][]{Hayes1957}. This has consequences for instabilities such as the Richtmyer-Meshkov instability which results in a hydrodynamic contact discontinuity being unstable, however the instability is suppressed in MHD \citep{Wheatley2009}. For the corrugation instability (where a shock front is perturbed), the opposite can be true where the instability grows in MHD but is suppressed in HD.

The stability of shocks to the corrugation instability depends on the type of shock. Fast MHD shocks are categorically stable provided the adiabatic index $\gamma <3$ \citep{Gardner1964}, whereas parallel MHD shocks 
(where the velocity and magnetic fields are parallel to the shock front)
are always unstable to the corrugation instability \citep{Stone1995}. The stability of other types of shocks (e.g., switch-off) depends on the angle of the magnetic field (relative to the shock front) and the Alfv\'en Mach number \citep{Stone1995,Edelman1989,Lessen1967}. Here we focus on a slow-mode (parallel) shock that is categorically unstable for MHD, and stable for HD.

The stability of shock fronts to perturbations is important for compressible turbulent astrophysical systems, such as the interstellar medium \citep{Elmegreen2004} and solar atmosphere \citep{Reardon2008}, where slow-mode shocks can encounter a non-uniform medium and develop instability. This may make slow-mode shocks more difficult to identify than fast mode shocks due to the corrugated shock front \citep{Park2019}. Similar results are present in reconnection studies that feature turbulence \citep{Zenitani2011,Zenitani2020}.

In addition to MHD and HD systems, shocks regularly occur in warm plasma, where the medium is only partially ionised, such as the solar chromosphere and molecular clouds. These partially ionised mediums represent a challenging area of study, specifically for instabilities, where the general behaviour is neither MHD-like or HD-like.
Partially-ionised mediums can be modelled using two-fluid (neutral, ion+electron) equations that allow for coupling and decoupling of the two species, and non-MHD-like behaviour. 
Two-fluid interactions in interface instabilities (where the initial magnetic field is parallel to the jump) can suppress small-scale features \citep{Popescu2020} and allow cross-field transport of momentum and energy \citep{Hillier2019}. 

The stability of two-fluid shocks to the corrugation instability has not been studied and likely depends on the level of coupling of the plasma and neutral species.
A consequence of partial ionisation on shocks is that the shock wave has a finite-width that is determined by the physical parameters of the system. \cite{Hillier2016} extensively studied the two-fluid effects in a switch-off slow-mode shock. Within the finite width of the shock additional shock transitions can occur due to the species decoupling \citep{Snow2019}. One can hypothesise that under the extreme conditions of the finite width being zero or infinity, the plasma should reduce to MHD-like behaviour. However, in the more realistic finitely-coupled regimes the behaviour is undetermined, which forms the basis of the study presented here.

In this paper, we study the stability of two-fluid slow-mode shock fronts using 2D numerical simulations. We investigate the consequences and stability of the shock front for different levels of collisional coupling and different ionisation fractions. We find that the shock-front can be stable or unstable depending on the wavelength of the perturbation relative to the finite width of the shock. This may have consequences for observations by giving an expected range at which partially-ionised slow-mode shocks can be stable.   

\begin{figure}
    \centering
    \includegraphics[width=0.95\linewidth,clip=true,trim=0.8cm 7.8cm 1.5cm 9.4cm]{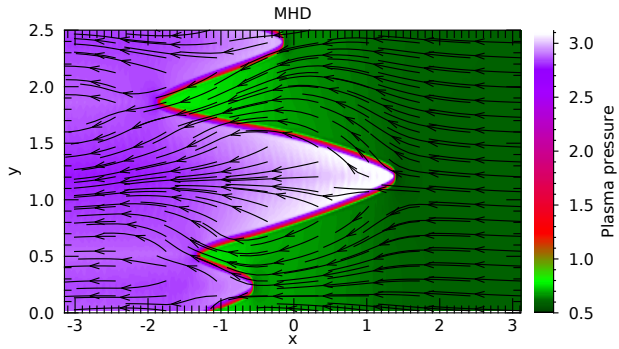}
    \caption{Magnetohydrodynamic (MHD) schematic for flow entering a distorted shock front. The colourmap shows the plasma pressure. Velocity vectors are overplotted, with the $v_y$ velocity multiplied by a factor of 15 to emphasise the flow}
    \label{fig:schematicMHD}
\end{figure}

\section{Corrugation instability schematic} \label{sec:corschem}

In this paper, we focus on specifically the parallel slow-mode shock, where the flow is aligned with the magnetic field either side of the shock. This type of shock is unconditionally unstable in the MHD model \citep{Stone1995}, and stable in the hydrodynamic model, to the corrugation instability. In this section, we present the basic schematic for the HD stability and MHD instability for a 2D shock front encountering an upstream density perturbation, and make conjectures about the potential stability or instability expected for a partially ionised system.  

\subsection{MHD}

An ideal magnetohydrodynamic (MHD) shock is unstable to the corrugation instability under certain conditions. 
For a parallel MHD shock, where the velocity and magnetic fields are parallel to the shock front either side of the shock (i.e., $\textbf{v}=[v_x,0,0],\textbf{B}=[B_x,0,0]$ in the shock frame), the shock front is always unstable to the corrugation instability \citep{Stone1995}.
The steady-state shock jump conditions can be written as
\begin{gather}
    \left[\rho v_x  \right]^u _d = 0, \label{eqn:jump1} \\
    \left[B_x \right]^u _d = 0, \\
    \rho v_x \left[v_y \right]^u _d = B_x \left[ B_y \right]^u _d, \label{eqn:sjumpmx} \\
    \rho v_x \left[\frac{B_y}{\rho} \right]^u _d = B_x \left[ v_y \right]^u _d, 
\end{gather}
\begin{gather}
    \left[\rho v_x^2 +P + \frac{1}{2} B_y^2 \right]^u _d = 0, \\
    \rho v_x \left[\frac{1}{2} (v_x^2 +v_y^2) + \frac{P}{(\gamma -1) \rho} +\frac{P}{\rho} +\frac{B_y^2}{\rho} \right]^u _d = B_x \left[ v_y \right]^u _d, \label{eqn:jump6}
\end{gather}
where the above notation relates conserved quantities upstream (superscript $u$) and downstream (superscript $d$) of the shock front as:
\begin{gather}
    \left[ Q \right]^u _d \equiv Q^u - Q^d
\end{gather}
for any conserved quantity $Q$.
From these equations (particularly Equation \ref{eqn:sjumpmx}) it can be seen that a jump in the tangential component of the velocity can be supported across a steady-state shock front by a jump in the tangential magnetic field component.

When the shock front corrugates it generates a tangential magnetic field component and hence a tangential velocity. This velocity is continuous across the interface and hence the flow is directed behind the peaks, see Figure\ref{fig:schematicMHD}.  
The flow pattern increases the pressure behind the peaks and decreases it behind the troughs. As such, the peaks are pushed up and the troughs are pushed down, increasing the magnitude of the corrugation and leading to growth of the instability. This is shown by the schematic in Figure \ref{fig:schematicMHD}.

\subsection{Hydrodynamics}

\begin{figure}
    \centering
    \includegraphics[width=0.95\linewidth,clip=true,trim=0.8cm 7.8cm 1.5cm 9.4cm]{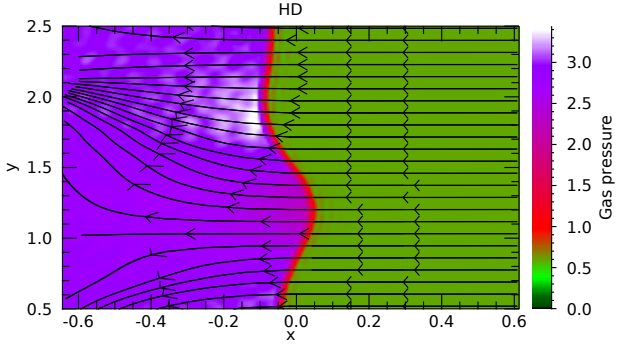}
    \caption{Hydrodynamic (HD) schematic for flow entering a distorted shock front. The colourmap shows the gas pressure. Velocity vectors are overplotted as the black lines, with the $v_y$ velocity multiplied by a factor of 30 to emphasise the flow.}
    \label{fig:schematicHD}
\end{figure}

An ideal hydrodynamic (HD) shock is stable to the corrugation instability. In a HD system, tangential velocity is continuous across the shock front. Therefore, at the distorted shock front the flow is refracted away from the 'peaks'.
The flow then enhances the pressure behind the 'troughs', and decreases the pressure behind the 'peaks', see Figure \ref{fig:schematicHD}. 
As such, the pressure provides a restoring force whereby the peaks are pushed down and the troughs are pushed up. The pressure restoring force therefore acts to mitigate the corrugation of the shock front and the system stabilises, returning to a flat shock front. 
A schematic of this is shown in Figure \ref{fig:schematicHD} where a high pressure region forms behind the troughs.

\subsection{Interpolating these results to partially-ionised plasma (PIP)}

In the two fluid partially ionised case, the schematic is not as straightforward. The ionised species behaves like an MHD system, whereas the neutral species is HD. 
As such, one may consider that the isolated neutral species tends towards stability, whereas the isolated plasma species tends towards being unstable.
The stability of the bulk (plasma+neutral) PIP shock depends on the balance of these forces, and the timescales on which the species interact.

The coupling between the ionised and neutral species is governed by the collisional coefficient $\alpha _c$ \citep{Hillier2016} which is discussed further in Section \ref{sec:methods}. In the the extreme of $\alpha _c =0$, the system is fully decoupled, and the ionised species will be unstable. At the other extreme of $\alpha_c=\infty$ the system is fully coupled and MHD-like, hence one would also expect the system to be unstable. In the finitely coupled regime, interactions can allow the system to be unstable or stable. The finitely coupled regime is the focus of our paper. 

Similarly, one has to consider the perturbation wavelength relative to the finite width of the shock. 
Consider a fixed size perturbation of wavelength $\lambda _{\parallel}$ that is parallel to the shock front, at different levels of coupling. For a infinitely coupled system, the finite width of the shock approaches zero and hence the perturbation wavelength is much larger than the finite width of the shock. As such, one may expect that the system will behave similar to a bulk-MHD model. At the other extreme of coupling tending to zero, the finite width of the shock becomes large. The upstream perturbation in this case will encounter the plasma shock front first and therefore also behave like an MHD simulation. Between these two extremes, we expect the system to be finitely coupled. As such, the stability of a partially-ionised shock can be connected to the wavelength of the perturbation relative to the finite shock width. 


\section{Methods} \label{sec:methods}

\begin{figure}
    \centering
    \includegraphics[width=0.95\linewidth,clip=true,trim=1.45cm 8.5cm 2.8cm 8.8cm]{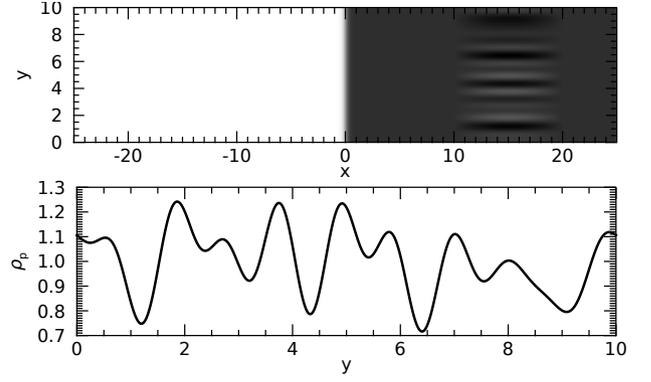}
    \caption{Initial density profile for the MHD simulation. Top panel shows a contour of the density. The shock is located at $x=0$. A density perturbation is upstream of the shock centred at $x=15$. A line plot of the density perturbation is shown in the bottom panel. }
    \label{fig:IC}
\end{figure}

The two-fluid equations governing the behaviour of a neutral species (subscript $\rm n$) and a charge-neutral electron+ion species (subscript $\rm p$) are:

\begin{gather}
\frac{\partial \rho _{\text{n}}}{\partial t} + \nabla \cdot (\rho _{\text{n}} \textbf{v}_{\text{n}})= 0, \label{eqn:neutral1} \\
\frac{\partial}{\partial t}(\rho _{\text{n}} \textbf{v}_{\text{n}}) + \nabla \cdot (\rho _{\text{n}} \textbf{v}_{\text{n}} \textbf{v}_{\text{n}} + P_{\text{n}} \textbf{I}) = -\alpha _c \rho_{\text{n}} \rho_{\text{p}} (\textbf{v}_{\text{n}}-\textbf{v}_{\text{p}}) \\
\frac{\partial e_{\text{n}}}{\partial t} + \nabla \cdot \left[\textbf{v}_{\text{n}} (e_{\text{n}} +P_{\text{n}}) \right] \nonumber \\  \hspace{0.5cm}= -\alpha _c \rho _{\text{n}} \rho _{\text{p}} \left[ \frac{1}{2} (\textbf{v}_{\text{n}} ^2 - \textbf{v}_{\text{p}} ^2)+ \frac{3}{2} \left(\frac{P_n}{\rho_n}-\frac{1}{2}\frac{P_p}{\rho_p}\right) \right],  \\
e_{\text{n}} = \frac{P_{\text{n}}}{\gamma -1} + \frac{1}{2} \rho _{\text{n}} v_{\text{n}} ^2, \label{eqn:neutral2}\\
\frac{\partial \rho _{\text{p}}}{\partial t} + \nabla \cdot (\rho_{\text{p}} \textbf{v}_{\text{p}}) = 0 \label{eqn:plasma1}\\
\frac{\partial}{\partial t} (\rho_{\text{p}} \textbf{v}_{\text{p}})+ \nabla \cdot \left( \rho_{\text{p}} \textbf{v}_{\text{p}} \textbf{v}_{\text{p}} + P_{\text{p}} \textbf{I} - \textbf{B B} + \frac{\textbf{B}^2}{2} \textbf{I} \right) \\
\frac{\partial}{\partial t} \left( e_{\text{p}} + \frac{\textbf{B}^2}{2} \right) + \nabla \cdot \left[ \textbf{v}_{\text{p}} ( e_{\text{p}} + P_{\text{p}}) -  (\textbf{v}_p \times \textbf{B}) \times \textbf{B} \right] \nonumber \\  \hspace{0.5cm} =  \alpha _c \rho _{\text{n}} \rho _{\text{p}} \left[ \frac{1}{2} (\textbf{v}_{\text{n}} ^2 - \textbf{v}_{\text{p}} ^2)+ \frac{3}{2} \left(\frac{P_n}{\rho_n}-\frac{1}{2}\frac{P_p}{\rho_p}\right) \right],\\
\frac{\partial \textbf{B}}{\partial t} - \nabla \times (\textbf{v}_{\text{p}} \times \textbf{B}) = 0, \\
e_{\text{p}} = \frac{P_{\text{p}}}{\gamma -1} + \frac{1}{2} \rho _{\text{p}} v_{\text{p}} ^2, \\
\nabla \cdot \textbf{B} = 0,\label{eqn:plasma2}
\end{gather}
for density $\rho$, energy $e$, pressure $P$, velocity $\textbf{v}$ and magnetic field $\textbf{B}$. The adiabatic index $\gamma=5/3$ and is constant. The plasma and neutral species are coupled through thermal collisions which is controlled through the coupling coefficient $\alpha _c$,
\begin{gather}
    \alpha_c = \alpha _0 \sqrt{\frac{T_n+T_p}{2}},
\end{gather}
where $T_n,T_p$ are the neutral and plasma temperatures respectively. The constant $\alpha _0$ can be specified and governs the coupling between the two species. We study a hydrogen-only system. 

We use a 4th-order central difference solver with artificial viscosity to limit the numerical oscillations around the shock front. It was found this solver is far less susceptible to the numerical Carbunkle instability \citep[which forms due to a curved feature forming on a Cartesian grid,][]{ELLING2009} than the first order HLLD solver used in previous shock studies with the (P\underline{I}P) code \citep{Snow2019,Snow2020,Snow2021}. The 4th order scheme in the (P\underline{I}P) code has been used successfully in prior studies of partially ionised plasma dynamics \citep{Hillier2019,Murtas2021}.

\subsection{Initial conditions}

The initial conditions used in this paper are in the shock frame, where the shock is stationary at $x=0$, with the inflow coming from the upstream conditions ($x>0$), and the post-shock region downstream ($x<0$). This has the advantage of allowing us to increase the resolution of the system since no space is required for the shock to propagate into.

A sonic Mach $2$ parallel shock is specified analytically, see Appendix \ref{app:par}, where `parallel' refers to the initial flow being aligned with the magnetic field. Here, the shock is in the $x$-direction and $v_y=B_y=0$ initially and corresponds to a slow-mode shock. The pressure and density are specified such that we have a thermal equilibrium and the initial plasma-$\beta=0.1$ in the upstream medium.

The corrugation is initiated by placing a small pseudo-random density perturbation upstream of the shock for the form
\begin{gather}
        \rho= 
\begin{cases}
    \rho_0 + \sum_{n=1}^{10} A \sin\left(\frac{\left(x-x_0\right) \pi}{(x_1-x_0)}\right) \cos \left(\frac{2\pi n (y-y_c)}{(y_1-y_0)}\right),& \text{if } x_0 \leq x\leq x_1\\
    \rho_0,              & \text{otherwise}
\end{cases}
\end{gather}
where $x_0,x_1,y_0,y_1$ give the horizontal and vertical locations for the perturbation. 
10 wavelengths are imposed parallel to the shock front (i.e., in the $y$ direction), with random amplitude $A$ and phase (determined by $y_c$).
The perturbation is placed sufficiently upstream of the shock that the shock front has time to relax to its numerical equilibrium before encountering the density enhancement. Note that the initial shock front is uniform in the $y-$direction so evolves like a 1D system before it encounters the density perturbation. We initially set the perturbation between $x_0=10,x_1=20$ and $y_0=0,y_1=10$ covering the full vertical extent of our box and horizontally located upstream of the shock front. The same random seed (and hence the same perturbation) is used in all simulations. The initial conditions are shown in Figure \ref{fig:IC} along with a 1D plot of the density perturbation. The perturbation is placed sufficiently far upstream to allow the shock front to relax slightly to its numerical equilibrium before the shock front encounters the density enhancement.

For the two-fluid case considered here (where the species are thermally coupled), the plasma conditions sufficiently upstream and downstream of the shock and described by the MHD shock jump equations \citep{Snow2019,Snow2021}. As such, the bulk (plasma+neutral) density, pressure, velocity and magnetic field jumps are the same as the MHD case, see Appendix \ref{app:par}. The partial pressures and densities are calculated using the neutral fraction and the bulk pressure and density. The neutral and plasma parameters are therefore defined as:
\begin{gather}
    \rho_p=(1-\xi_n) \rho_t, \\
    \rho_n=\xi_n \rho_t, \\
    v_{px}=v_{nx}=v_{tx}, \\
    v_{py}=v_{ny}=v_{ty}, \\
    P_p= \frac{2\xi_p}{\xi_n+2\xi_p} P_t,\\
    P_n= \frac{\xi_n}{\xi_n+2\xi_p} P_t,
\end{gather}
where a subscript $t$ denotes the MHD value and $\xi_n$ and $\xi_p$ are the neutral and plasma fractions respectively. In this formation, the MHD simulation is effectively the infinitely coupled case where the system behaves like a bulk (plasma+neutral) fluid. At the other extreme of $\alpha_c=0$ the fluids are completely decoupled and the plasma will behave MHD-like, free of any interactions with the neutral species. The shock front is allowed to numerically stabilise before encountering the upstream density perturbation.

\begin{figure}
    \centering
\includegraphics[width=0.95\linewidth,clip=true,trim=4.88cm 4.8cm 5.15cm 3.85cm]{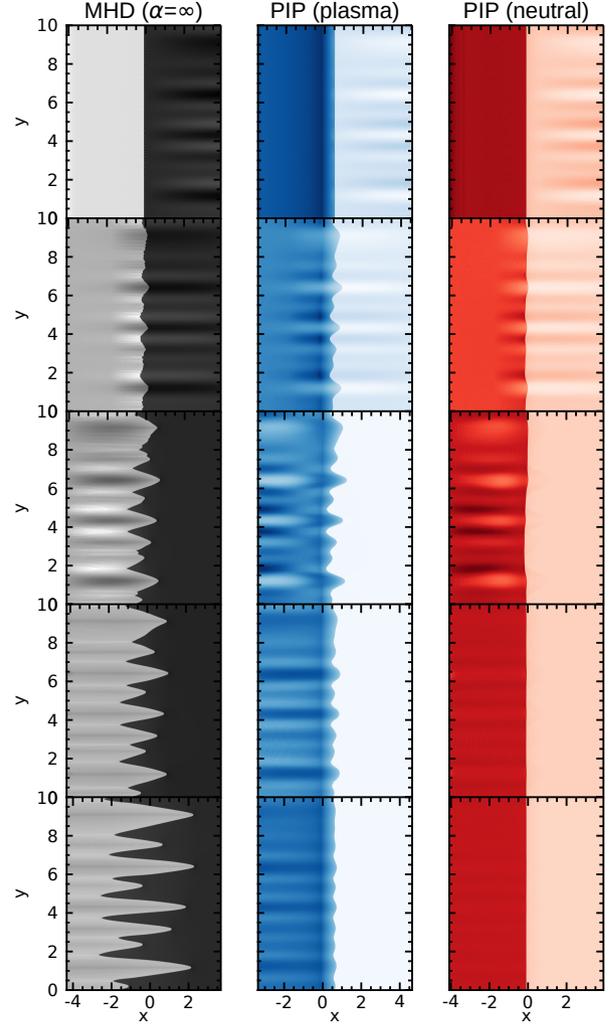}
    \caption{Time series showing the evolution of the density around the shock front in the MHD (left), and PIP (plasma centre, neutral right). Snapshots are taken at times  $t=5,7,10,15,30$ non-dimensional times from top to bottom. }
    \label{fig:timeseries}
\end{figure}

\subsection{Boundary conditions}

Upper and lower $(y=0,10)$ boundary conditions are set to periodic. The left and right boundaries ($x=-25,25$) are specified as damping layers using a similar formulation as \cite{Felipe2010}. Several waves are generated from the initial shock wave relaxing and the interactions with the density perturbation. This damping layer effectively removes these unwanted waves from the system.
\begin{gather}
    \textbf{U} = \textbf{U}_0 +\left(1-D(x)\right) \left(\textbf{U}-\textbf{U}_0 \right) \\
    D(x)=\frac{D_0}{\Delta x} \left( \frac{x-x_{D}}{W}\right)^2
\end{gather}
where $\textbf{U}$ is the instantaneous evolved vector of conserved variables and $\textbf{U}_0$ is the equilibrium quantity. The initial quantities $\textbf{U}_0$ are determined analytically using the shock jump conditions. The perturbation to be damped is then $\textbf{U}-\textbf{U}_0$ which is gradually damped over $W=20$ grid cells. $x_D$ is the $x$ value at the end of the damping region. The factor $D_0$ scales the maximum damping to be 0.3. We find that this boundary condition is sufficient to remove unwanted perturbations from system and prevents reflections from the boundary. The boundaries are also placed far from the shock front to further minimise their impact on the results.  

\section{Reference model}

\subsection{MHD simulation}

\begin{figure}
    \centering
    \includegraphics[width=0.95\linewidth,clip=true,trim=1.1cm 7.8cm 1.5cm 7.8cm]{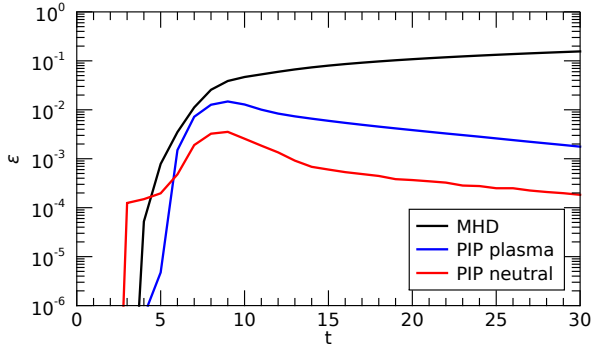}
    \caption{Integrated enstrophy around the shock front for the fiducial MHD (black) and PIP ($\xi_n =0.9,\alpha_c=1$) (blue plasma, red neutrals) models through time.}
    \label{fig:growthfid}
\end{figure}

\begin{figure}
    \centering
    \includegraphics[width=0.95\linewidth,clip=true,trim=1.2cm 7.8cm 1.3cm 7.8cm]{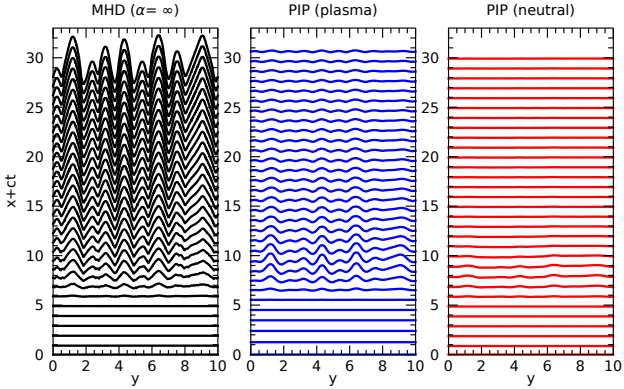}
    \caption{Evolution of the MHD (left) and PIP (plasma centre, neutral right) shock front displacement in time. The 1D shock front is extracted using the maximum density gradient in the $x$-direction. The constant $c$ is a unit speed.}
    \label{fig:shockevomhd}
\end{figure}

For the MHD system, the shock front encounters the density perturbation, corrugates, then the shock front perturbation grows with time, see Figure \ref{fig:timeseries}. The initial perturbation of the shock front allows a non-zero $B_y$ field to exist at the shock interface leading to growth of the instability, as described in Section \ref{sec:corschem}. 

In this paper, we quantify the corrugation using the integrated enstrophy (vorticity squared) around the shock front as:
\begin{gather}
    \varepsilon=\int _{0} ^{10} \int _{x_s-4} ^{x_s+4} \left( \nabla \times \textbf{v} \right) ^2 dx dy,
\end{gather}
where $x_s$ is the location of the shock front. The integral limits capture the full extend of the domain in the $y$-direction, and the horizontal window is chosen to capture the full extent of the corrugation in the $x$-direction. The vorticity is non-zero at the distorted shock front, and zero elsewhere. As such, the integrated enstrophy allows us to determine if the corrugation is growing or decaying with time. 

For the MHD model, Figure \ref{fig:growthfid}, the integrated enstrophy around the shock is close to zero initially, as it should be for a uniform shock front. When the shock front encounters the density enhancement, there is a sudden rise in the enstrophy as the shock front corrugates. As time increase, the corrugation continues to grow, as seen in Figure \ref{fig:timeseries}, and hence the integrated enstrophy grows. As time increases towards infinity the growth of the instability will saturate, as seen in \cite{Stone1995}, however this has not yet occurred in time frame studied here. 

A 1D shock front can be extracted from the 2D simulation by taking the displacement at each height $y$, with the location of the shock determined by the maximum density gradient in the $x$-direction. 
Plotting the shock front through time shows the growth of the corrugation, as shown in Figure \ref{fig:shockevomhd}. The initial disturbance warps the shock front. As time increases, the displacement grows across all frequencies. This shows the expected result that the MHD simulation is unstable to the corrugation instability.

\subsection{PIP simulation} \label{sec:pipfid}

\begin{figure}
    \centering
    \includegraphics[width=0.95\linewidth,clip=true,trim=0.8cm 8.0cm 1.2cm 7.8cm]{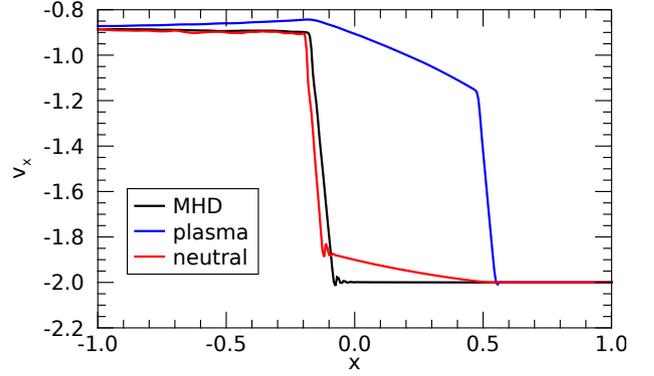}
    \caption{Initial shock for the MHD (black) and PIP (red neutrals, blue plasma) at $t=6$. The snapshot is taken before the shock front collides with the density enhancement. Small numerical oscillations exist at the shock front as a consequence of the numerical scheme.}
    \label{fig:inishock}
\end{figure}

\begin{figure}
    \centering
    \includegraphics[width=0.95\linewidth,clip=true,trim=0.8cm 7.9cm 1.5cm 9.4cm]{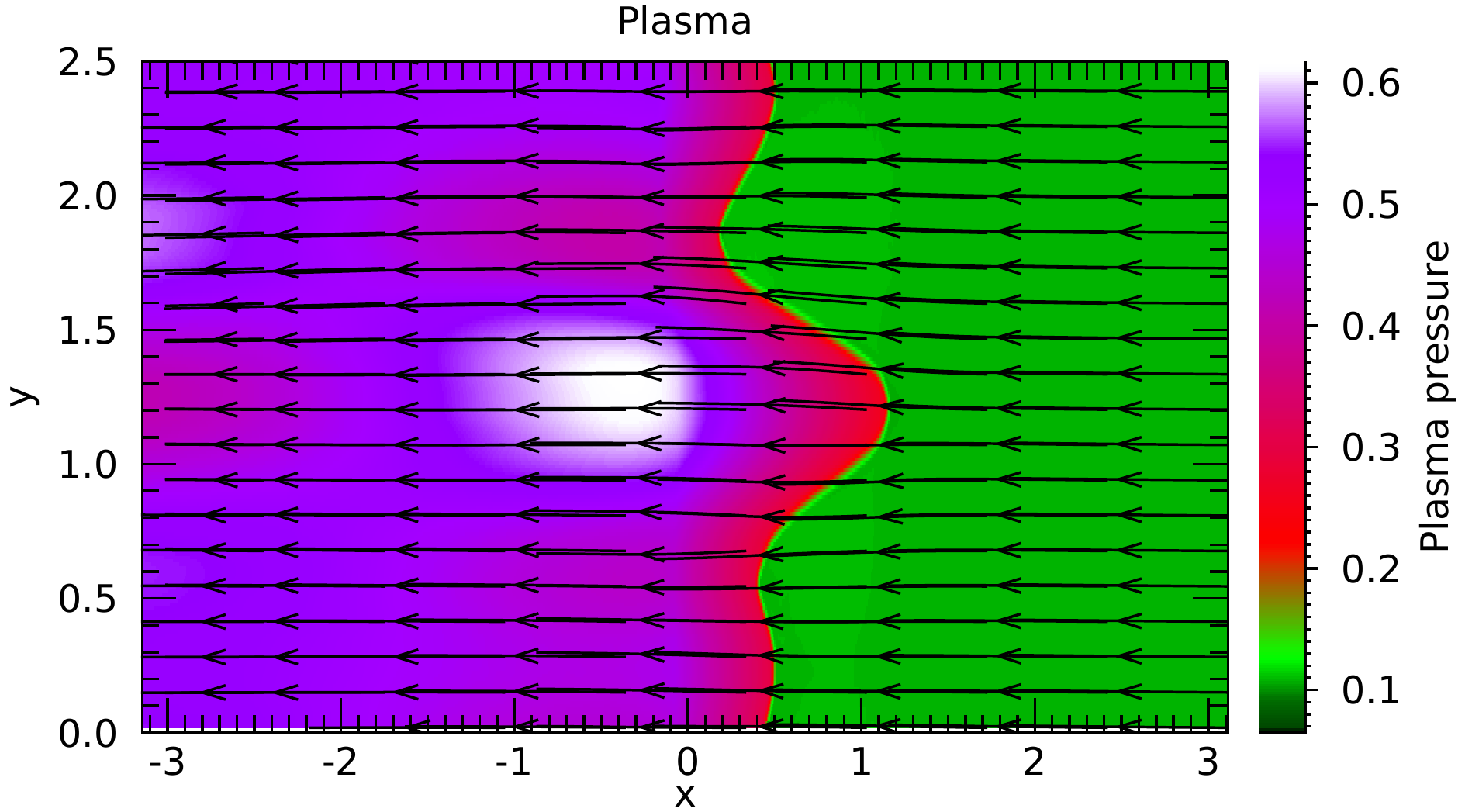} \\
    \includegraphics[width=0.95\linewidth,clip=true,trim=0.8cm 7.9cm 1.5cm 9.4cm]{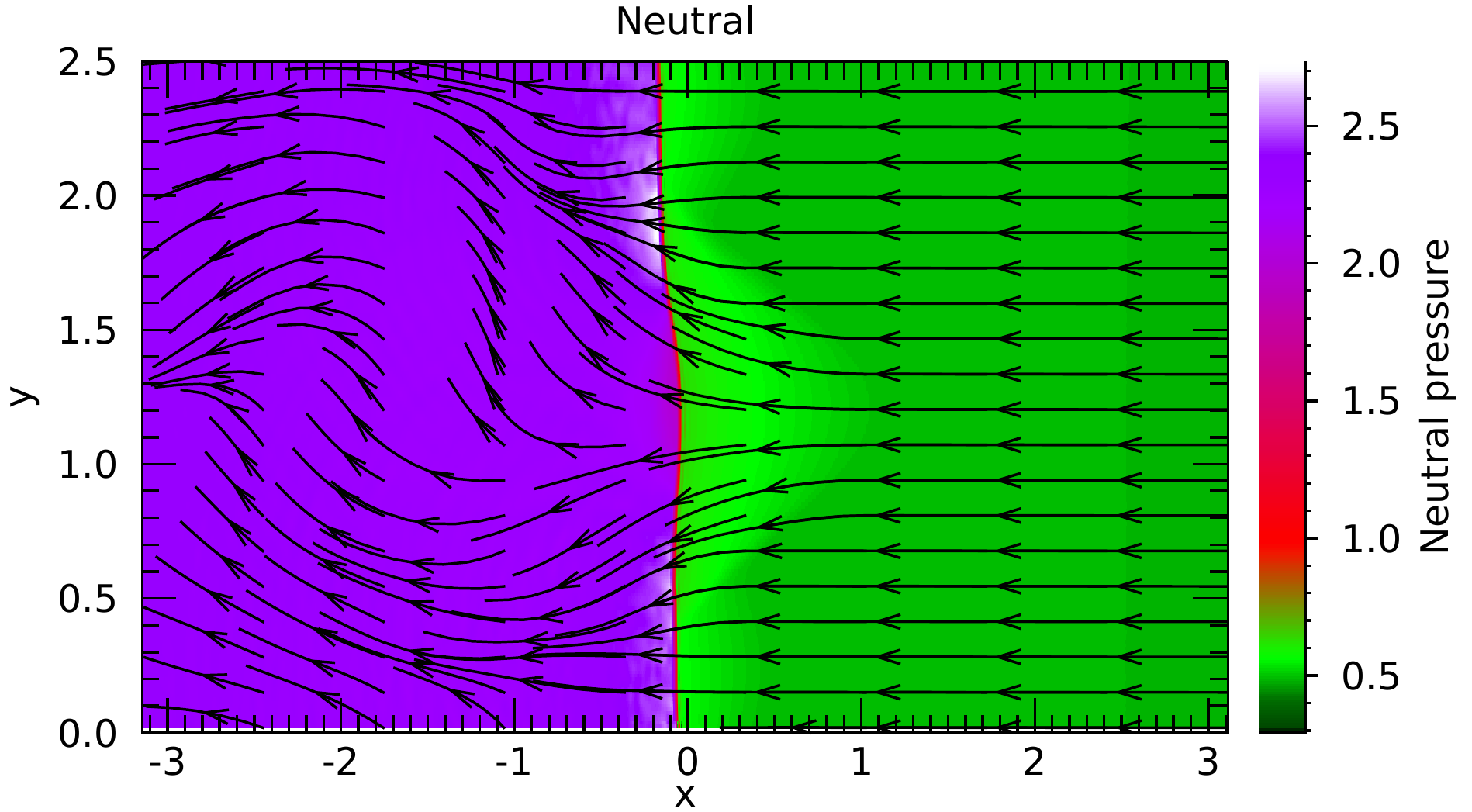}
    \caption{PIP flow schematic for the plasma and neutral species. The colour shows the species pressure. Velocity vectors are overplotted as arrows. The $v_y$ velocity is multiplied by a factor of 15 for visualisation purposes.}
    \label{fig:schemPIPxin09}
\end{figure}

\begin{figure}
    \centering
    \includegraphics[width=0.95\linewidth,clip=true,trim=0.8cm 8.2cm 1.1cm 8.0cm]{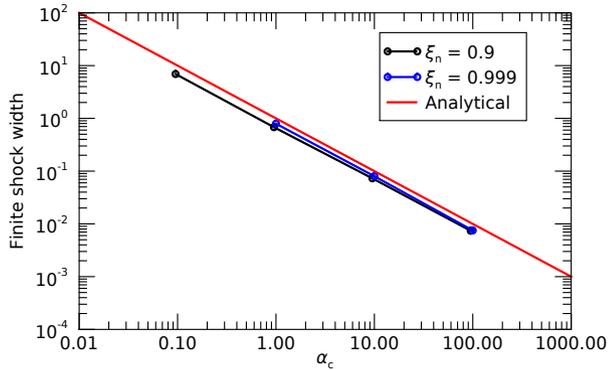}
    \caption{Shock width as a function of collisional coefficient $\alpha_c$ for the analytic expression (Equation \ref{eqn:shockwidth}) and 1D numerical simulations in the absence of perturbations.}
    \label{fig:shockwidth}
\end{figure}

\begin{figure*}
    \centering
\includegraphics[width=0.9\linewidth,clip=true,trim=1.8cm 8.0cm 2.0cm 6.5cm]{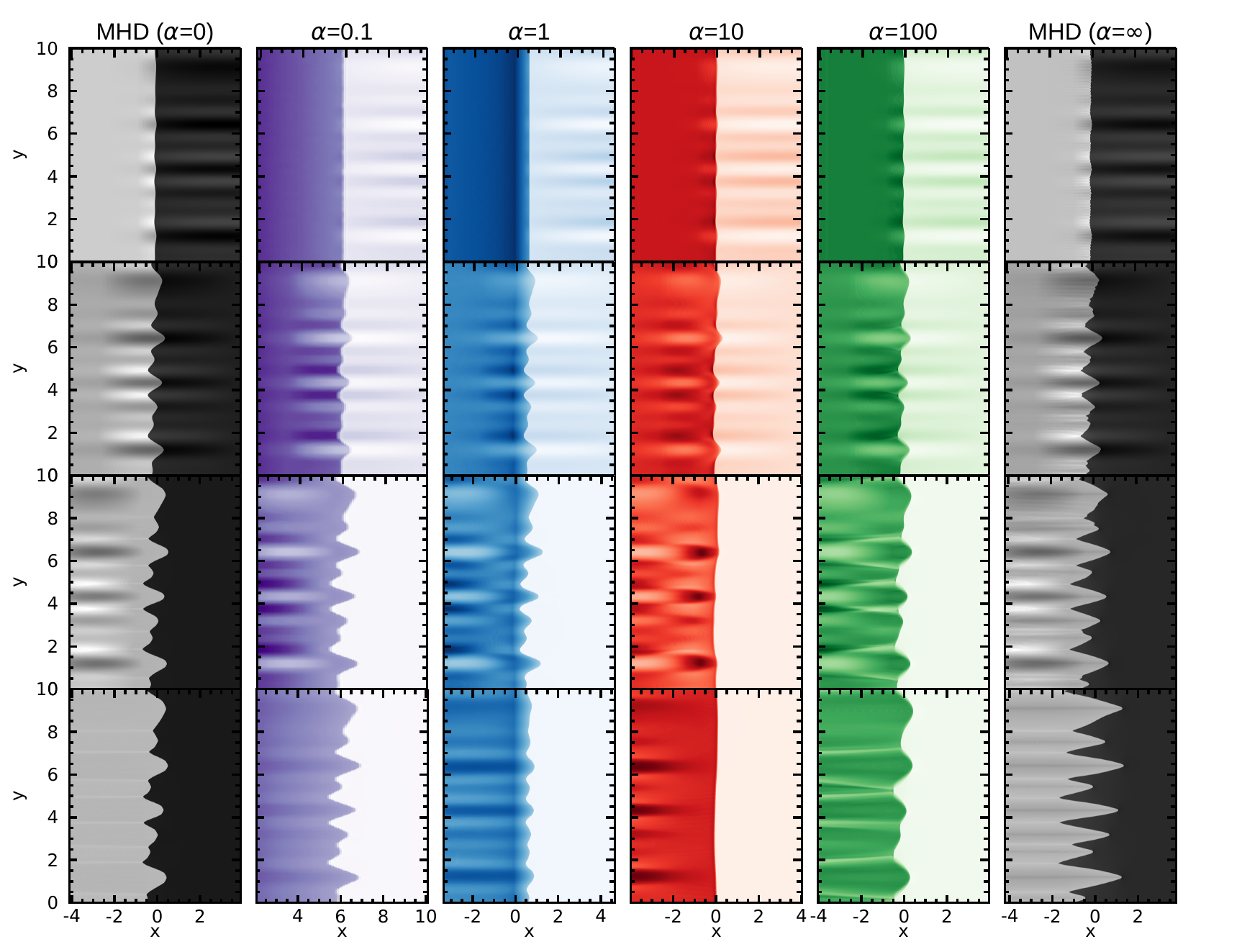}
    \caption{Time series of the shock front for different levels of collisional coupling and two MHD simulations for the extremes of zero and infinite coupling.}
    \label{fig:timeseriesac}
\end{figure*}

Here we present an initial two-fluid PIP simulation where the neutral fraction is set as $\xi_n=0.9$ and a coupling coefficient of $\alpha_0=1$, giving a mostly neutral medium that is relatively weakly coupled on the time scales considered here. This choice of coupling coefficent and initial conditions means that on average, upstream collisions occur on a time scale of unity. The bulk (neutral+plasma) density and pressure are the same as in the MHD case, see Appendix \ref{app:par}. 
For a two-fluid system, the bulk (neutral+plasma) shock jumps reduce to the MHD shock jumps sufficiently upstream and downstream of the shock
\citep{Snow2019}. However, within the finite-width of the shock, drift velocities form between the two species and substructure exists. The role of this substructure on the stability of the shock front is investigated here.

Figure \ref{fig:timeseries} shows the plasma and neutral densities through time for this fiducial PIP simulation. Initially the shock front is uniform in the $y-$direction, as with the MHD case, however the two-fluid effects produce a finite width to the shock which is highlighted in Figure \ref{fig:inishock} (the finite width of two-fluid shocks is discussed in detail in \cite{Hillier2016}). When the shock front encounters the density enhancement, corrugations occur in both the plasma and neutral shock fronts. These corrugations initially grow in the plasma and decay in the neutrals. Finally, the distortions decay in both species and the shock front stabilises. This can be seen in Figure \ref{fig:growthfid}, where there is a rise in corrugation, followed by a decay towards a stable shock front.

For this simulation, the plasma shock is at the leading edge, followed by the neutral shock, see Figure \ref{fig:inishock}. As such, the plasma species encounters the density perturbation first, and the instability grows in an MHD-like manner initially. At this time, there is also a slight corrugation of the neutral shock front from the encountered perturbation. However, as time advances, the displacement of the shock front decreases in both the neutral and plasma species and the shock front stabilises.

The schematic for flow entering the initially distorted shock is given in Figure \ref{fig:schemPIPxin09} and can be compared to the schematics for MHD (Figure \ref{fig:schematicMHD}) and HD (Figure \ref{fig:schematicHD}).
The flow entering the plasma shock is slightly corrugated in the same way as the MHD case. The $v_{py}$ magnitude is far less in the PIP case than the MHD case due to the resistance provided by the neutral species.
In the neutral species, the flow is directed towards the troughs, creating a high pressure region that suppresses the instability. 
In this regime with a mostly neutral medium ($\xi_n=0.9$) and a weakly coupled system ($\alpha _0 =1$) the interactions between the plasma and neutral species are sufficient to suppress the corrugation instability and stabilise the shock front.
This is further shown in the evolution of the shock front displacement with time, Figure \ref{fig:shockevomhd}, where the plasma shock front distorts initially, then stabilises. The neutral shock front distorts slightly after encountering the density enhancement but is quickly stabilised.
Physically, the neutral species is trying to stabilise the shock front, whereas the plasma species is trying to go unstable. Here the overall physics is dominated neutral species since medium is mostly neutral (hence the plasma-neutral coupling is prevalent) and the perturbation wavelengths in the $y-$direction are equally affected by the coupling (this is discussed further in Section \ref{sec:parstud}).

The finite width of the shock front here is $\approx 0.6$, which is smaller than parallel perturbation wavelengths that are in the range of $\lambda _{\parallel}=1-10$ nondimensional units. The ratio of the shock width to the perturbation wavelength should govern the behaviour of the system. For a perturbation much larger than the finite width, the system should behave like a fully coupled MHD simulation since the finite-shock width becomes negligible and overall behaviour is determined by bulk properties. For a perturbation much smaller than the finite width, the simulation should behave like a fully decoupled simulation, also MHD like, since on short time scales the perturbation will only encounter the plasma shock front. Between these two limits, the finite coupling leads to non-MHD-like behaviour.

\section{Parameter study} \label{sec:parstud}

\subsection{Coupling coefficient}

Changing the coupling coefficient changes the finite width of the shock, with larger coupling leading to a narrower finite width \citep{Hillier2016}. 
For the two-fluid $M=2$ shock investigated in this paper, with an upstream plasma-$\beta=0.1$, the finite width of the shock is a function of the coupling coefficient, as shown in Figure \ref{fig:shockwidth} from a series of 1D numerical simulations.
An analytic estimate is also shown, which is calculated as the difference in propagation speeds of the isolated fluids, divided by the coupling frequency, i.e.,
\begin{gather}
    W=\frac{c_{ps}-c_{ns}}{\alpha_c \rho}, \label{eqn:shockwidth}
\end{gather}
for the plasma and neutral sound speeds $c_{ps}$ and $c_{ns}$. 
The analytical and numerical results for the shock width pair up very well, and scale well in the coupling regime investigated here. Also, changing the neutral fraction does not greatly affect the shock width and this estimate remains a good fit to the 1D simulation data. 

At the limit of $\alpha_c = \infty$ and $\alpha_c = 0$, the system is MHD with the fluid acting as a bulk (neutral+plasma) system or a completely decoupled system respectively. Between these limits, we expect two-fluid effects to become important, as seen in Section \ref{sec:pipfid}. Here we investigate the consequences of changing the coupling coefficient on the stability of the shock front.

\begin{figure}
    \centering
    \includegraphics[width=0.95\linewidth,clip=true,trim=0.7cm 7.8cm 1.3cm 7.8cm]{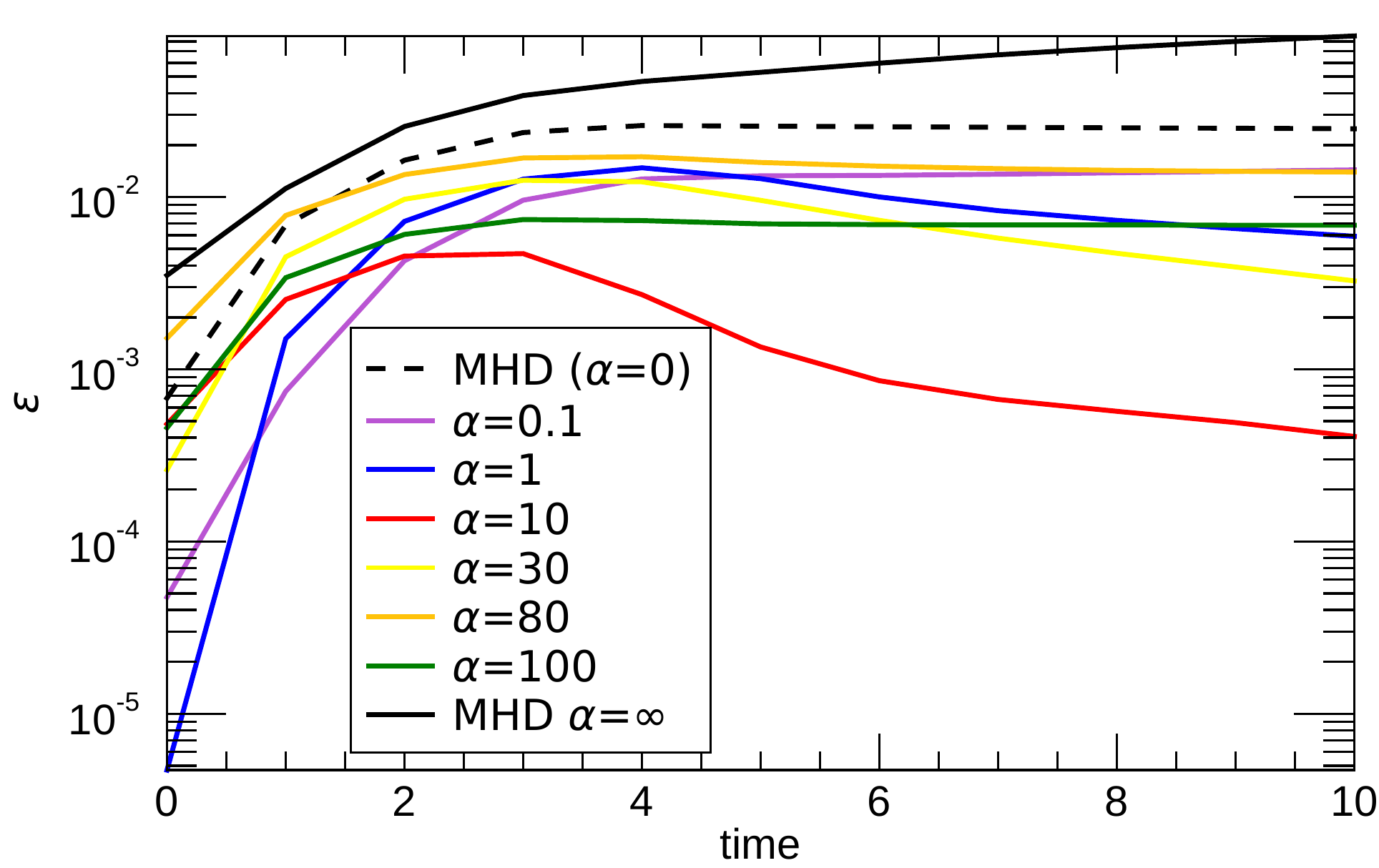} \\
    \includegraphics[width=0.95\linewidth,clip=true,trim=0.8cm 7.8cm 1.05cm 7.8cm]{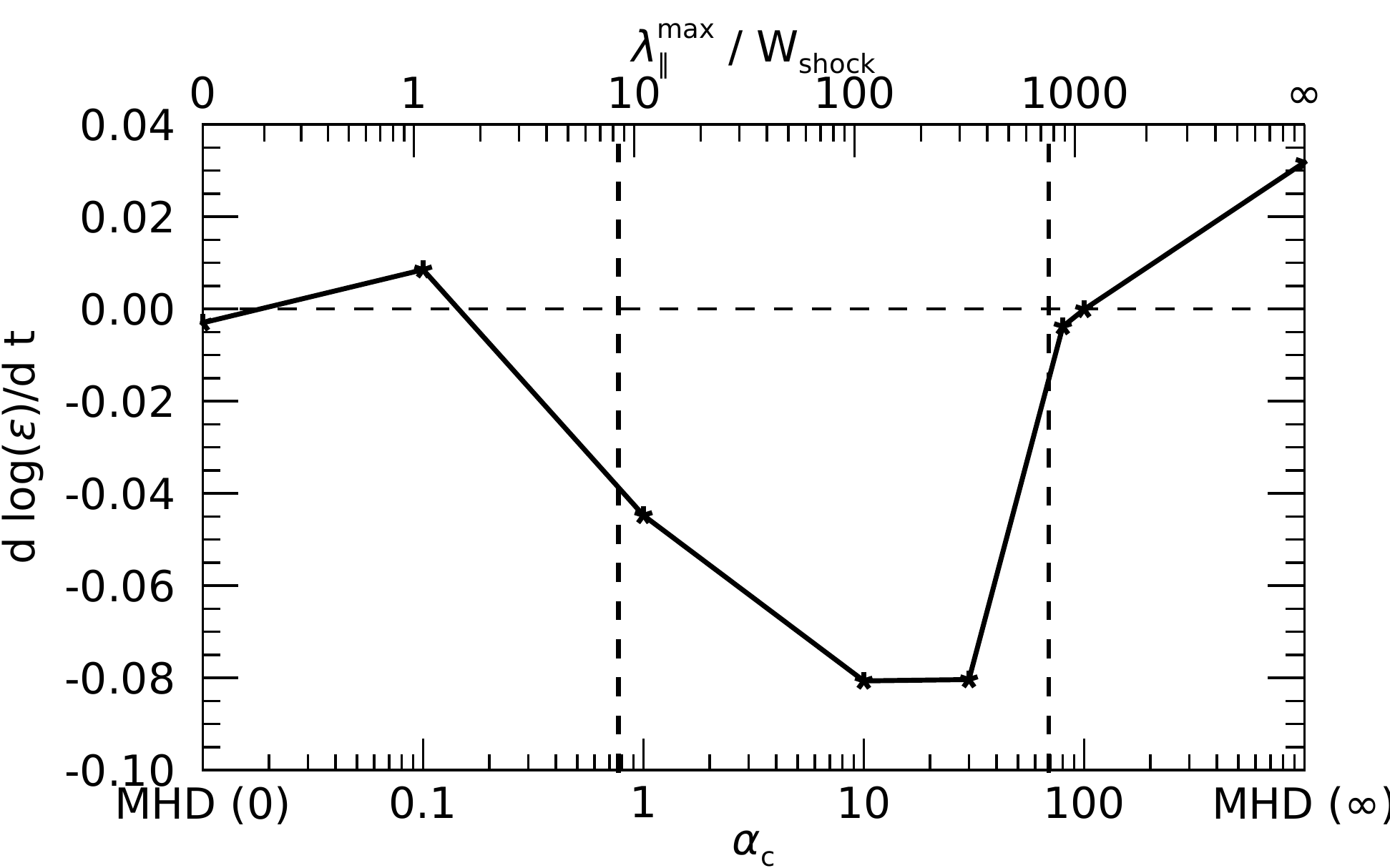}
    \caption{(Top) Growth rates for different levels of collisional coupling through time. Time zero corresponds to the perturbation encountering the shock front. (Lower) Time derivative of the log enstrophy corresponding to time 10 in the top figure, where a value greater than zero states growth of the instability, whereas a negative value states a stabilising shock front. The MHD simulation for $\alpha _c=0$ has stagnated growth and hence the value is close to zero. 
    A second axis is included for the largest parallel wavelength $\lambda _\parallel ^{\mbox{max}} =10$ divided by the finite width of the shock $W_{\mbox{shock}}$. The vertical dashed lines indicate the theoretical stability range based on Equations \ref{eqn:acmin}-\ref{eqn:acmax}. }
    \label{fig:growthac}
\end{figure}

Figure \ref{fig:timeseriesac} shows a time evolution of the shock front for different levels of coupling, and two MHD simulations that represent the infinitely coupled and fully decoupled extremes. Figure \ref{fig:growthac} shows the corresponding integrated entrophy around the shock front through time for these cases. One can see that the MHD simulations, the corrugation grows with time and persist throughout the simulation. The growth of the $\alpha =0$ case saturates near the end of the simulation, whereas the $\alpha=\infty$ case continues to grow. In both cases however, the shock front has become unstable and the corrugation is a consistent feature of the system. In the PIP simulations, the initial perturbation distorts the shock front (as with the MHD cases) however the perturbation is seen to stabilise to varying degrees depending on the coupling coefficient. For the $\alpha=1$ case, all frequencies are damped at roughly the same rate, with the shock front tending to a flat shock front. For $\alpha=10$ the moderate coupling has a severe effect in rapidly damping the perturbation, quickly leading to a stable/flat shock front. At $\alpha =100$ the perturbation grows and has saturated, however the smaller frequencies have been damped from the system and only larger frequencies exist.   

The integrated enstrophy around the shock is shown in Figure \ref{fig:growthac}a where the time zero corresponds to the shock encountering the perturbation. The corresponding time derivative at time $\tau=10$ is shown in Figure \ref{fig:growthac}b. The MHD ($\alpha _c=0$) saturates during the simulation and hence the time derivative is very close to zero, and in fact slightly negative. It can be seen from Figure \ref{fig:timeseriesac} that the perturbed shock front remains distorted with time and the growth rate being close to zero is due to saturation of the instability. For a weakly coupled system ($\alpha _c=0.1$) the system is unstable and growing, however the rate is very small implying that it is approaching saturation. At the end time, the simulation is comparable to the MHD ($\alpha_c=0$) case. The perturbation width for $\alpha_c=0.1$ is approximately the same size as the finite width of the shock. As the coupling increases ($\alpha_c=1,10$), the system becomes stable as the neutrals start to influence the plasma. This is seen clearly in Figure \ref{fig:timeseriesac} where these cases approach an undisturbed state at late times. At strong coupling ($\alpha _c =100$), the system begins to act like an ensemble bulk MHD ($\alpha_c=\infty$) system and the perturbation persists through time. From Figure \ref{fig:growthac}b it can be seen that the growth rate here is very close to zero and likely saturated for $\alpha_c=100$. The structure of the perturbed shock front in this case is also slightly different to the MHD $\alpha=\infty$ case; only the low frequency components are present in the perturbation, and it is much smoother than the MHD case.    

\begin{figure}
    \centering
    \includegraphics[width=0.95\linewidth,clip=true,trim=0.8cm 8.4cm 1.3cm 7.8cm]{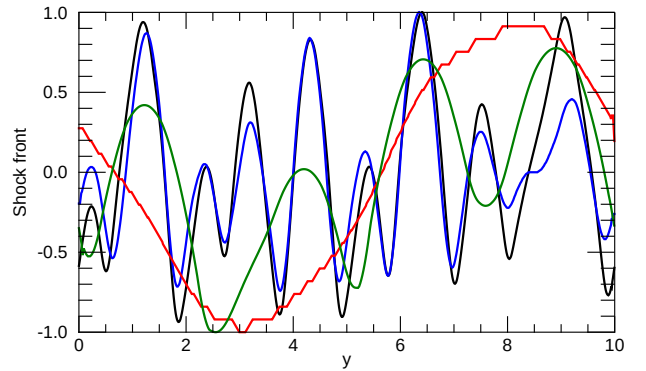}
    \caption{Normalised shock front for different coupling coefficients at time $t=20$. The maximum amplitude in each case is artificially scaled to unity for visualisation purposes.}
    \label{fig:shockfrontcomp}
\end{figure}

The wavelengths that persist in the shock front vary for different coupling coefficients, see Figure \ref{fig:shockfrontcomp} which shows the normalised shock front displacement. For weak coupling ($\alpha _c =1$) the wavelengths are all damped fairly uniformly, as seen in Figure \ref{fig:shockevomhd}. For moderate coupling ($\alpha_c =10$) the system stabilises very rapidly and only the longest wavelengths persists. This is shown in Figure \ref{fig:shockfrontcomp} where only the fundamental mode is present. For strong coupling ($\alpha_c =100$), the shorter scales are damped, and several longer scales persist. It is known that there are several scales on which the neutral and plasma species interact \citep{Hillier2019}. On the largest scales, the neutrals and plasma species are coupled. On intermediate length scales the neutrals are decoupled from the plasma, and on the smallest length scales both species are decoupled from each other. 
These scales are determined using the collision coupling frequencies:
\begin{gather}
 \nu _{in} = \rho _{tot} \xi _n \alpha _c, \\
 \nu _{ni} = \rho _{tot} (1-\xi _n) \alpha _c,
\end{gather}
where $\nu_{in}$ and $\nu_{ni}$ are the ion-neutral and neutral-ion collisional frequencies respectively. As $\alpha _c$ increases, the coupling frequencies increase. One can separate the coupling with wavelength into three frequency bands. For low frequencies, the neutrals and plasma are well coupled. For intermediate frequencies, the neutrals decouple from the plasma (and hence the magnetic field), however the plasma is still coupled to the neutrals. For high frequencies, both species decouple. For the simulations performed here, increasing the coupling coefficient effectively decreases the wavelength at which the neutrals decouple from the plasma. This can be used to explain the damping of short wavelengths in the $\alpha_c=100$ case.

The approximate coupling frequency at which the system becomes unstable can be estimated by equating the coupling frequency to the frequency of the wavelength. For the lower limit, the frequency at which the plasma decouples from the neutrals for the largest wavelength is
\begin{gather}
    \alpha_{c,\mbox{min}}  \rho ^u _n = 2 \pi \frac{|v^u-v^d|}{\lambda _\parallel ^{\mbox{max}}}. \label{eqn:acmin}
\end{gather}
Similarly, the upper limit is provided by the frequency at which the neutral decoupled from the plasma using the smallest wavelength:
\begin{gather}
    \alpha_{c,\mbox{max}}  \rho ^u _p = 2 \pi \frac{|v^u-v^d|}{\lambda _\parallel ^{\mbox{min}}}. \label{eqn:acmax}
\end{gather}
The velocity jump in our simulation is $|v^u-v^d| \approx 1.1$, the upstream densities are $\rho ^u _p = 0.1$,$\rho ^u _n = 0.9$, and our wavelengths are in the range $\lambda _\parallel \in \mathbb{Z} : \lambda _\parallel \in [1,10]$, with $\lambda _\parallel ^{\mbox{max}}=10$ and $\lambda _\parallel ^{\mbox{min}}=1$. 
As such, we estimate that the upper and lower limits on stability of our system are $\alpha_{c,\mbox{min}} \approx 0.754$ and $\alpha_{c,\mbox{max}} \approx 69.1$, which are shown by the dashed lines on Figure \ref{fig:growthac}. 
One can see that these approximations are a reasonably good fit to the simulation results, and all simulations between $\alpha_{c,\mbox{min}} < \alpha _c < \alpha_{c,\mbox{max}}$ are stable to the corrugation instability.

We note that a time scale argument exists for the weakly coupled systems.
On the time scales considered here the corrugation evolves and reaches saturation, however, as time advances, the plasma will experience more collisions from the neutral species. As such, as time tends to infinity, the saturated, corrugated PIP shocks may further evolve due to a gradual influence of the neutral species.

\begin{figure*}
    \centering
    \includegraphics[width=0.95\linewidth,clip=true,trim=2.0cm 8.0cm 2.5cm 6.7cm]{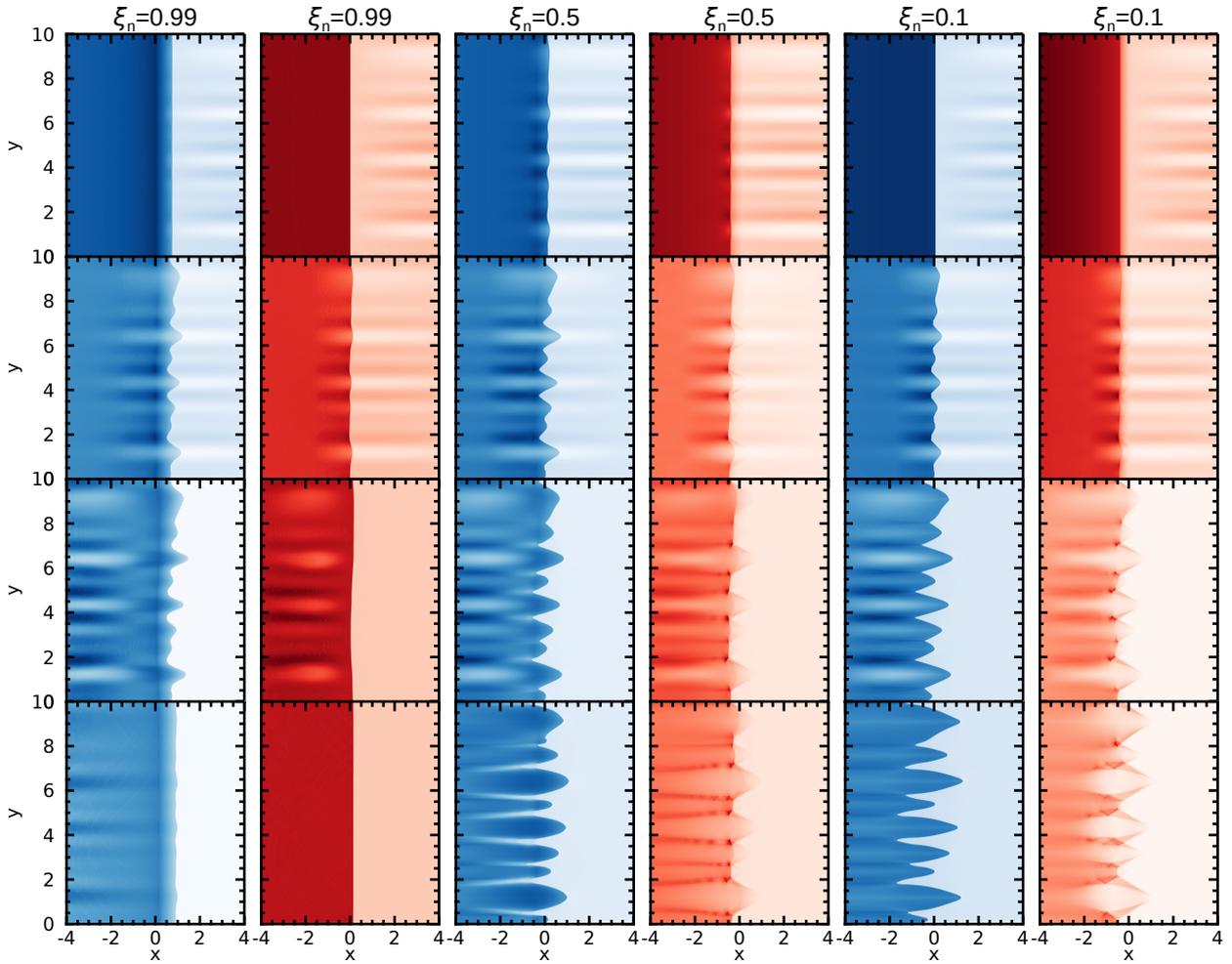}
    \caption{Time series of the plasma (blue) and neutral (red) densities for different initial neutral fractions. Snapshots are taken at $\tau=0,2,5,12$, where $\tau=0$ is when the shock front first encounters the upstream density perturbation.}
    \label{fig:timeseriescompxin}
\end{figure*}

\begin{figure}
    \centering
    \includegraphics[width=0.95\linewidth,clip=true,trim=0.8cm 7.9cm 1.5cm 9.4cm]{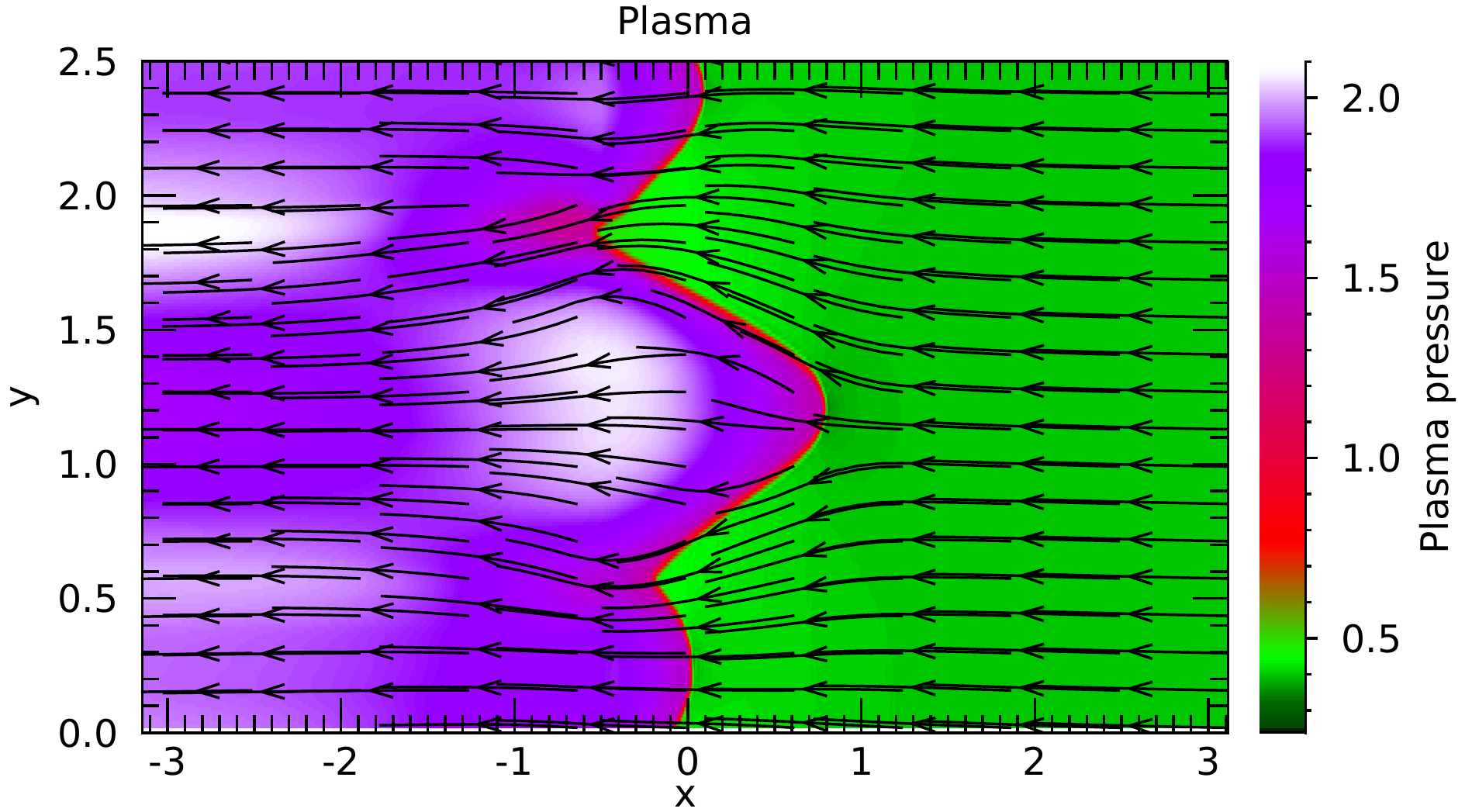} \\
    \includegraphics[width=0.95\linewidth,clip=true,trim=0.8cm 7.9cm 1.5cm 9.4cm]{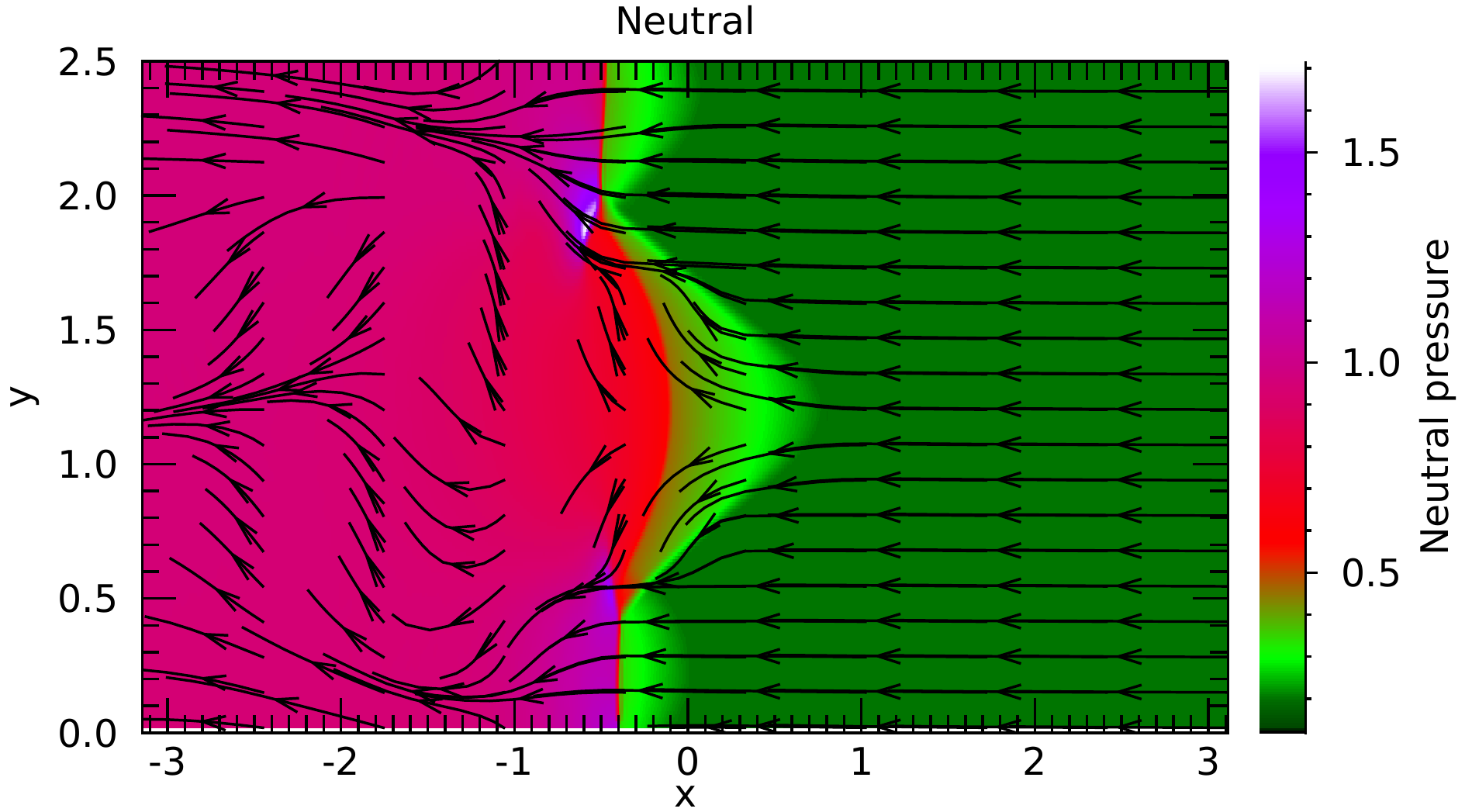}
    \caption{PIP flow schematic for the plasma (top) and neutral (bottom) species for the $\xi_n=0.5$ case at $t=10$. The colour shows the species pressure. Velocity vectors are overplotted as arrows. The $v_y$ velocity is multiplied by a factor of 15 for visualisation purposes.}
    \label{fig:schemPIPxin05}
\end{figure}

\subsection{Ionisation fraction}

Simulations thus far used a neutral fraction of $\xi _n=0.9$ meaning that the medium is mostly neutral particles. Here we investigate changing the neutral fraction by looking at $\xi_n=0.99,0.5,0.1$. The $\xi_n=0.1$ case is mostly ionised. The $\xi_n=0.99$ case is dominated by the neutral species. The $\xi_n=0.5$ is an interesting case where the system is equally ionised and neutral. A time series showing the plasma and neutral densities at different times is shown in Figure \ref{fig:timeseriescompxin}.

The $\xi_n=0.99$ case behaves fairly similar to the $\xi_n=0.9$ case studied previously. Both the plasma and neutral shock fronts corrugate slightly when they encounters the upstream density perturbation. As time advances, the shock front stabilises. Since this simulation is dominated by the neutral species, the stabilising forces from the neutral species are stronger than the destabilising forces of the plasma. The corrugation of the plasma shock front increases between times $t=2,5$. However the neutral coupling then leads to a stabilisation of the shock front. This is comparable to the results of the $\xi_n=0.9$ case in Section \ref{sec:pipfid}.

For a neutral fraction of $\xi _n =0.5$ the system is equally neutral and ionised. The evolution of the shock front through time is shown in Figure \ref{fig:timeseriescompxin}. The shock front is unstable to the corrugation instability here and the shock front distortion grows with time. Interestingly, this simulation forms far longer fingers than has been seen in the MHD or PIP simulations in Section \ref{sec:pipfid}
Between the elongated fingers, the neutral species also develops shock channels, as seen by the sharp density structures. 

In the MHD case, the flow is directed behind the peaks (see Figure \ref{fig:schematicMHD}). For the PIP $\xi_n=0.5$ case, the neutral species resists this pattern and the plasma flow is directed towards the troughs instead, see Figure \ref{fig:schemPIPxin05}. As time advances, this forces the troughs to extend further back and creates the elongated fingers seen in the Figure \ref{fig:timeseriescompxin}. In the neutral species, similar behaviour occurs where the flow is directed towards the troughs, resulting in an increased neutral pressure and the formation of shocks due to the converging flow, Figure \ref{fig:schemPIPxin05}. When the plasma can exert a strong influence on the neutrals, these shock channels can form as flow pattern of the neutral species is dominated by the plasma motion.

The $\xi_n=0.1$ case is dominated by the plasma species, and, as such, the corrugation of the plasma shock front grows, similar to the MHD model. The behaviour of the neutral species is mostly determined through the coupling with the plasma species. The result is that the neutral flow is channelled towards the troughs, as with the $\xi_n=0.5$ case. However, here the neutral species constitutes far less of the bulk medium and hence there is very little feedback from the neutrals to the plasma.


\section{Discussion}

\subsection{Perturbation length scale}

One can re-frame the effect of different collisional coefficients in terms of the length scale relative to finite width of the shock. The initial case studied in Section \ref{sec:pipfid} has a finite shock width of approximately $W=0.68$ and an initial parallel perturbation wavelengths in the range $\lambda _\parallel =[1,10]$, therefore the wavelength of the perturbation is larger than the finite width of the shock.
The perturbation wavelength relative to the finite width of the shock is an important parameter. The simulations in Section \ref{sec:parstud} showed that when the perturbation wavelength was between 10 and 1000 times larger than the finite width, the neutral interactions stabilised the shock front.

The finite-width of the parallel slow shock considered here can be estimated using Equation \ref{eqn:shockwidth}. Rewriting this in terms of bulk fluid properties:
\begin{gather}
    c_{sp}^2= \frac{\gamma P_p}{\rho_p} =\frac{\gamma P_B}{\rho_B} \frac{\chi _i}{\xi_i} \\
    c_{sn}^2= \frac{\gamma P_n}{\rho_n} =\frac{\gamma P_B}{\rho_B} \frac{\chi _n}{\xi_n} \\
    \chi_i= \frac{2 \xi_i}{\xi_n-2\xi_i} \\
    \chi_n= \frac{\xi_n}{\xi_n-2\xi_i} \\
    c_{sp}-c_{sn}= \sqrt{\frac{\gamma P_B}{\rho _B}} \left( \sqrt{\frac{\chi_i}{\xi_i}} - \sqrt{\frac{\chi_n}{\xi_n}} \right) 
\end{gather}
The bulk sound speed in the chromosphere is approximately $8$ km/s. The ion-neutral collisional frequency in the chromosphere varies in the range of $10^{3}-10^{6}$\,s$^{-1}$ \citep{Popescu2019}. This leads to a finite-width of approximately $\approx 4$ m. In Section \ref{sec:parstud} it is shown that perturbations 1000 times larger than finite width were unstable. Therefore, we expect that in the partially ionised solar chromosphere, slow-mode shock fronts will be stable to perturbations on the order of approximately $4-4000$ m. Physically, this means that the perturbations that we are able to observe should result in unstable shock fronts. Below our observational limits, the partially ionised shocks should stabilise. 

The analytical approximation for the stability range of a partially ionised parallel shock (given by Equation \ref{eqn:acmin}-\ref{eqn:acmax}) is a reasonable approximation to the simulation results. This formulation can be written for a parallel slow-mode shock of an arbitrary sonic Mach number $M$ as:
\begin{gather}
    \alpha_{c,\mbox{min}}  \rho ^u _n= \frac{2 \pi c_s^u}{\lambda _\parallel ^{\mbox{max}}} \frac{M}{r}(r-1),\label{eqn:abrshock1} \\
    \alpha_{c,\mbox{max}}  \rho ^u _p= \frac{2 \pi c_s^u}{\lambda _\parallel ^{\mbox{min}}} \frac{M}{r}(r-1) \label{eqn:abrshock2},
\end{gather}
for compressible ratio $r=\rho^d/\rho^u$, where $r>1$ for a shock. Using the relationship between $r$ and $M$ for a parallel slow-mode shock:
\begin{gather}
    r=\frac{M^2(\gamma+1)}{2+M^2(\gamma -1)} 
\end{gather}
we can rewrite Equations \ref{eqn:abrshock1},\ref{eqn:abrshock2} as
\begin{gather}
    \alpha_{c,\mbox{min}}  \rho ^u _n= \frac{2 \pi c_s^u}{\lambda _\parallel ^{\mbox{max}}} \frac{2(M^2-1)}{M(\gamma -1)},\label{eqn:abrshock3} \\
    \alpha_{c,\mbox{max}}  \rho ^u _p= \frac{2 \pi c_s^u}{\lambda _\parallel ^{\mbox{min}}} \frac{2(M^2-1)}{M(\gamma -1)} \label{eqn:abrshock4}.
\end{gather}
Therefore, given the upstream plasma and neutral densities, the upstream sound speed, the shock Mach number, and the range of perturbation wavelengths upstream of the shock, one can estimate the coupling frequencies at which the the shock may become unstable. These coupling frequencies can then be compared to the expected range of frequencies for the medium \citep[e.g., ][]{Popescu2019}

The results can be used to approximate the stability of shocks in a solar sunspot, where the magnetic field is fairly straight and propagating shocks are regularly observed in the form of umbral flashes \citep{Beckers1969}, which have average lifetimes of around 44.2 seconds \citep{Nelson2017}. Umbral flashes have Mach numbers between $M=1$ and $M=1.7$ \citep{Anan2019} and using the coupling frequencies in the upper chromosphere \citep[e.g., see Figure 2 in][]{Popescu2019} of $\nu_{in}=3\times 10^2,\nu_{ni}=3$ s$^{-1}$ with a typical sound speed of $c_s=8$ km/s gives a stable range of wavelengths between approximately 0.6 and 56 km (for M=1.7).

For the parallel slow-mode shock investigated here, the shock width has very little dependence on the ionisation fraction, see Figure \ref{fig:shockwidth}. However, it is known that for switch-off shocks, the finite width has a dependence on the neutral fraction and plasma$-\beta$ \citep{Hillier2016}. Also, the physical width of the shocks studied in this paper is much smaller than the $>300$\,km shocks in \cite{Snow2020} As such, one would expect that the stability range changes for different shock types. Further work is needed to calculate this stable range for different types of partially-ionised shocks.

\subsection{Slow-mode shocks in turbulence}

\begin{figure}
    \centering
    \includegraphics[width=0.95\linewidth,clip=true,trim=1.0cm 8.0cm 2.5cm 8.0cm]{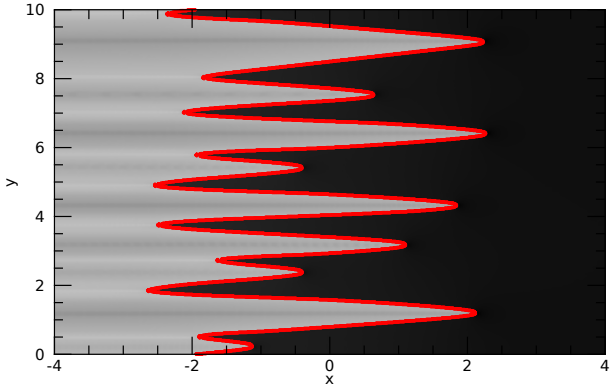}
    \caption{Detected pixels that satisfy a slow-mode shock transition in the reference MHD simulation at time $t=30$. Note that the detected shock pixels cover the majority of the shock front.}
    \label{fig:shockdetect}
\end{figure}

In MHD simulations of turbulence, fast-mode shocks are identified more readily than slow-mode shocks \citep{Park2019}. The potential reason for this is that the slow-mode shocks are unstable to the corrugation instability, hence the shock front deforms and is difficult to identify. However, analysing the MHD simulation of the corrugation instability by using an automated shock detection method \citep[based on the SHOCKFIND algorithm developed by][]{Lehmann2016} we see that the slow-mode shocks are still detected over the vast majority of the shock front, see Figure \ref{fig:shockdetect}. In face, due to the increased length of the shock, approximately 10 times more points are detected in the corrugated shock front at time $t=30$ than in the initial stable shock front. This implies that the corrugation instability should lead to a greater detection of slow-mode shocks, contrary to the claim of \cite{Park2019}. Further study is needed to determine the effect of the corrugation instability on the presence of shocks for different levels of corrugation and Mach numbers.

If the corrugation instability is reducing the detection of slow mode shocks \citep[as suggested by][]{Park2019}, then in a partially ionised system, we may detect statistically more slow-mode shocks in a two-fluid turbulence than MHD turbulence. Here we show that for a partially ionised medium, the neutral interactions can stabilise slow-mode shocks. As such, the corrugation of the shock front is a transient feature and shocks should damp the perturbation.

\subsection{Extension to 3D shocks}

\cite{Stone1995} show that a wider range of shock types are unstable in 3D MHD, since perturbation are allowed to grow in the out-of-plane direction and MHD shocks that are stable in 2D can be unstable in 3D. Here we see that for an unstable 2D shock in a partially-ionised plasma, the neutral species can act to stabilise the system. In 3D we would expect that the neutral species could act to stabilise the corrugation growth in the out-of-plane direction, resulting in a stable 3D partially ionised shock. A future study will investigate the stability of 3D partially ionised shocks. 

\section{Conclusions}

In this paper we have investigated the stability of slow-mode shocks in partially ionised plasma to the corrugation instability. As the shock front encounters a density perturbation, the shock front distorts leading to the corrugation instability. For an MHD system, a parallel slow-mode shock is always unstable, whereas a HD shock is usually stable. For the partially-ionised cases presented here, we show that the instability can be stable or unstable depending on the system parameters.

In the reference case ($\alpha_c=1,\xi_n=0.9$), the corrugation grows in the plasma species initially, then the coupling with the neutral species leads to stabilisation of the shock front. The stability of the shock front is dependent on the coupling coefficient. At large coupling ($\alpha_c>>100$) the system is unstable to the corrugation instability and behaves like a bulk MHD system. At the other extreme ($\alpha<0.1$), the system is again unstable and MHD-like, behaving like a fully decoupled plasma. Between these limits ($0.1<\alpha_c<100$) the neutral interactions with the plasma species lead to a stabilisation of the shock front to the corrugation instability.   

The stability range can be rewritten in therms of the perturbation width relative to the finite-width of the shock, and determines the stability of the shock front. When the perturbation width is on the same order (or smaller) than the finite-width of the shock, the system is unstable and MHD-like. Similarly, when the perturbation is 1000 times larger than the finite-width, the system is again unstable and MHD-like. Between these limits, the two-fluid interactions become important and the neutral species can stabilise the shock front. 

In conclusion, two-fluid interactions can lead to stabilisation of MHD parallel slow-mode shock fronts to the corrugation instability. Further work is needed to determine the stability criteria of different shock types.  

\section*{Acknowledgements}

BS and AH are supported by STFC research grant ST/R000891/1. 
AH also acknowledges support by STFC Ernest Rutherford Fellowship grant number ST/L00397X/2

\section*{Data Availability}

The simulation data from this study are available from BS upon reasonable request. The (P\underline{I}P) code is available at \href{https://github.com/AstroSnow/PIP}{https://github.com/AstroSnow/PIP}.



\bibliographystyle{mnras}
\bibliography{bib} 

\begin{thebibliography}{}
\makeatletter
\relax
\def\mn@urlcharsother{\let\do\@makeother \do\$\do\&\do\#\do\^\do\_\do\%\do\~}
\def\mn@doi{\begingroup\mn@urlcharsother \@ifnextchar [ {\mn@doi@}
  {\mn@doi@[]}}
\def\mn@doi@[#1]#2{\def\@tempa{#1}\ifx\@tempa\@empty \href
  {http://dx.doi.org/#2} {doi:#2}\else \href {http://dx.doi.org/#2} {#1}\fi
  \endgroup}
\def\mn@eprint#1#2{\mn@eprint@#1:#2::\@nil}
\def\mn@eprint@arXiv#1{\href {http://arxiv.org/abs/#1} {{\tt arXiv:#1}}}
\def\mn@eprint@dblp#1{\href {http://dblp.uni-trier.de/rec/bibtex/#1.xml}
  {dblp:#1}}
\def\mn@eprint@#1:#2:#3:#4\@nil{\def\@tempa {#1}\def\@tempb {#2}\def\@tempc
  {#3}\ifx \@tempc \@empty \let \@tempc \@tempb \let \@tempb \@tempa \fi \ifx
  \@tempb \@empty \def\@tempb {arXiv}\fi \@ifundefined
  {mn@eprint@\@tempb}{\@tempb:\@tempc}{\expandafter \expandafter \csname
  mn@eprint@\@tempb\endcsname \expandafter{\@tempc}}}

\bibitem[\protect\citeauthoryear{{Anan}, {Schad}, {Jaeggli}  \& {Tarr}}{{Anan}
  et~al.}{2019}]{Anan2019}
{Anan} T.,  {Schad} T.~A.,  {Jaeggli} S.~A.,   {Tarr} L.~A.,  2019, \mn@doi
  [\apj] {10.3847/1538-4357/ab357f}, \href
  {https://ui.adsabs.harvard.edu/abs/2019ApJ...882..161A} {882, 161}

\bibitem[\protect\citeauthoryear{{Beckers} \& {Tallant}}{{Beckers} \&
  {Tallant}}{1969}]{Beckers1969}
{Beckers} J.~M.,  {Tallant} P.~E.,  1969, \mn@doi [\solphys]
  {10.1007/BF00146140}, \href
  {https://ui.adsabs.harvard.edu/abs/1969SoPh....7..351B} {7, 351}

\bibitem[\protect\citeauthoryear{{Draine}, {Roberge}  \& {Dalgarno}}{{Draine}
  et~al.}{1983}]{Draine1983}
{Draine} B.~T.,  {Roberge} W.~G.,   {Dalgarno} A.,  1983, \mn@doi [\apj]
  {10.1086/160617}, \href
  {https://ui.adsabs.harvard.edu/abs/1983ApJ...264..485D} {264, 485}

\bibitem[\protect\citeauthoryear{{{\'E}del'Man}}{{{\'E}del'Man}}{1989}]{Edelman1989}
{{\'E}del'Man} M.~A.,  1989, \mn@doi [Astrophysics] {10.1007/BF01006842}, \href
  {https://ui.adsabs.harvard.edu/abs/1989Ap.....31..656E} {31, 656}

\bibitem[\protect\citeauthoryear{Elling}{Elling}{2009}]{ELLING2009}
Elling V.,  2009, \mn@doi [Acta Mathematica Scientia]
  {https://doi.org/10.1016/S0252-9602(10)60007-0}, 29, 1647

\bibitem[\protect\citeauthoryear{{Elmegreen} \& {Scalo}}{{Elmegreen} \&
  {Scalo}}{2004}]{Elmegreen2004}
{Elmegreen} B.~G.,  {Scalo} J.,  2004, \mn@doi [\araa]
  {10.1146/annurev.astro.41.011802.094859}, \href
  {https://ui.adsabs.harvard.edu/abs/2004ARA&A..42..211E} {42, 211}

\bibitem[\protect\citeauthoryear{{Felipe}, {Khomenko}  \& {Collados}}{{Felipe}
  et~al.}{2010}]{Felipe2010}
{Felipe} T.,  {Khomenko} E.,   {Collados} M.,  2010, \mn@doi [\apj]
  {10.1088/0004-637X/719/1/357}, \href
  {https://ui.adsabs.harvard.edu/abs/2010ApJ...719..357F} {719, 357}

\bibitem[\protect\citeauthoryear{{Gardner} \& {Kruskal}}{{Gardner} \&
  {Kruskal}}{1964}]{Gardner1964}
{Gardner} C.~S.,  {Kruskal} M.~D.,  1964, \mn@doi [Physics of Fluids]
  {10.1063/1.1711271}, \href
  {https://ui.adsabs.harvard.edu/abs/1964PhFl....7..700G} {7, 700}

\bibitem[\protect\citeauthoryear{{Hayes}}{{Hayes}}{1957}]{Hayes1957}
{Hayes} W.~D.,  1957, \mn@doi [Journal of Fluid Mechanics]
  {10.1017/S0022112057000403}, \href
  {https://ui.adsabs.harvard.edu/abs/1957JFM.....2..595H} {2, 595}

\bibitem[\protect\citeauthoryear{{Hillier}}{{Hillier}}{2019}]{Hillier2019}
{Hillier} A.,  2019, \mn@doi [Physics of Plasmas] {10.1063/1.5103248}, \href
  {https://ui.adsabs.harvard.edu/abs/2019PhPl...26h2902H} {26, 082902}

\bibitem[\protect\citeauthoryear{{Hillier}, {Takasao}  \& {Nakamura}}{{Hillier}
  et~al.}{2016}]{Hillier2016}
{Hillier} A.,  {Takasao} S.,   {Nakamura} N.,  2016, \mn@doi [\aap]
  {10.1051/0004-6361/201628215}, \href
  {https://ui.adsabs.harvard.edu/abs/2016A&A...591A.112H} {591, A112}

\bibitem[\protect\citeauthoryear{{Hollweg}}{{Hollweg}}{1982}]{Hollweg1982}
{Hollweg} J.~V.,  1982, \mn@doi [\apj] {10.1086/159993}, \href
  {https://ui.adsabs.harvard.edu/abs/1982ApJ...257..345H} {257, 345}

\bibitem[\protect\citeauthoryear{{Lehmann}, {Federrath}  \& {Wardle}}{{Lehmann}
  et~al.}{2016}]{Lehmann2016}
{Lehmann} A.,  {Federrath} C.,   {Wardle} M.,  2016, \mn@doi [\mnras]
  {10.1093/mnras/stw2015}, \href
  {https://ui.adsabs.harvard.edu/abs/2016MNRAS.463.1026L} {463, 1026}

\bibitem[\protect\citeauthoryear{{Lessen} \& {Deshpande}}{{Lessen} \&
  {Deshpande}}{1967}]{Lessen1967}
{Lessen} M.,  {Deshpande} N.~V.,  1967, \mn@doi [Journal of Plasma Physics]
  {10.1017/S0022377800003457}, \href
  {https://ui.adsabs.harvard.edu/abs/1967JPlPh...1..463L} {1, 463}

\bibitem[\protect\citeauthoryear{{Murtas}, {Hillier}  \& {Snow}}{{Murtas}
  et~al.}{2021}]{Murtas2021}
{Murtas} G.,  {Hillier} A.,   {Snow} B.,  2021, arXiv e-prints, \href
  {https://ui.adsabs.harvard.edu/abs/2021arXiv210201630M} {p. arXiv:2102.01630}

\bibitem[\protect\citeauthoryear{{Nelson}, {Henriques}, {Mathioudakis}  \&
  {Keenan}}{{Nelson} et~al.}{2017}]{Nelson2017}
{Nelson} C.~J.,  {Henriques} V.~M.~J.,  {Mathioudakis} M.,   {Keenan} F.~P.,
  2017, \mn@doi [\aap] {10.1051/0004-6361/201730467}, \href
  {https://ui.adsabs.harvard.edu/abs/2017A&A...605A..14N} {605, A14}

\bibitem[\protect\citeauthoryear{{Park} \& {Ryu}}{{Park} \&
  {Ryu}}{2019}]{Park2019}
{Park} J.,  {Ryu} D.,  2019, \mn@doi [\apj] {10.3847/1538-4357/ab0d7e}, \href
  {https://ui.adsabs.harvard.edu/abs/2019ApJ...875....2P} {875, 2}

\bibitem[\protect\citeauthoryear{{Petschek}}{{Petschek}}{1964}]{Petschek1964}
{Petschek} H.~E.,  1964, {Magnetic Field Annihilation}.
p.~425

\bibitem[\protect\citeauthoryear{{Popescu Braileanu}, {Lukin}, {Khomenko}  \&
  {de Vicente}}{{Popescu Braileanu} et~al.}{2019}]{Popescu2019}
{Popescu Braileanu} B.,  {Lukin} V.~S.,  {Khomenko} E.,   {de Vicente} {\'A}.,
  2019, \mn@doi [\aap] {10.1051/0004-6361/201834154}, \href
  {https://ui.adsabs.harvard.edu/abs/2019A&A...627A..25P} {627, A25}

\bibitem[\protect\citeauthoryear{{Popescu Braileanu}, {Lukin}, {Khomenko}  \&
  {de Vicente}}{{Popescu Braileanu} et~al.}{2020}]{Popescu2020}
{Popescu Braileanu} B.,  {Lukin} V.~S.,  {Khomenko} E.,   {de Vicente} A.,
  2020, arXiv e-prints, \href
  {https://ui.adsabs.harvard.edu/abs/2020arXiv200715984P} {p. arXiv:2007.15984}

\bibitem[\protect\citeauthoryear{{Reardon}, {Lepreti}, {Carbone}  \&
  {Vecchio}}{{Reardon} et~al.}{2008}]{Reardon2008}
{Reardon} K.~P.,  {Lepreti} F.,  {Carbone} V.,   {Vecchio} A.,  2008, \mn@doi
  [\apjl] {10.1086/591790}, \href
  {https://ui.adsabs.harvard.edu/abs/2008ApJ...683L.207R} {683, L207}

\bibitem[\protect\citeauthoryear{{Snow} \& {Hillier}}{{Snow} \&
  {Hillier}}{2019}]{Snow2019}
{Snow} B.,  {Hillier} A.,  2019, \mn@doi [\aap] {10.1051/0004-6361/201935326},
  \href {https://ui.adsabs.harvard.edu/abs/2019A&A...626A..46S} {626, A46}

\bibitem[\protect\citeauthoryear{{Snow} \& {Hillier}}{{Snow} \&
  {Hillier}}{2020}]{Snow2020}
{Snow} B.,  {Hillier} A.,  2020, \mn@doi [\aap] {10.1051/0004-6361/202037848},
  \href {https://ui.adsabs.harvard.edu/abs/2020A&A...637A..97S} {637, A97}

\bibitem[\protect\citeauthoryear{{Snow} \& {Hillier}}{{Snow} \&
  {Hillier}}{2021}]{Snow2021}
{Snow} B.,  {Hillier} A.,  2021, \mn@doi [\aap] {10.1051/0004-6361/202039667},
  \href {https://ui.adsabs.harvard.edu/abs/2021A&A...645A..81S} {645, A81}

\bibitem[\protect\citeauthoryear{{Stone} \& {Edelman}}{{Stone} \&
  {Edelman}}{1995}]{Stone1995}
{Stone} J.~M.,  {Edelman} M.,  1995, \mn@doi [\apj] {10.1086/176476}, \href
  {https://ui.adsabs.harvard.edu/abs/1995ApJ...454..182S} {454, 182}

\bibitem[\protect\citeauthoryear{{Suematsu}, {Shibata}, {Neshikawa}  \&
  {Kitai}}{{Suematsu} et~al.}{1982}]{Suematsu1982}
{Suematsu} Y.,  {Shibata} K.,  {Neshikawa} T.,   {Kitai} R.,  1982, \mn@doi
  [\solphys] {10.1007/BF00153464}, \href
  {https://ui.adsabs.harvard.edu/abs/1982SoPh...75...99S} {75, 99}

\bibitem[\protect\citeauthoryear{{Tidman} \& {Krall}}{{Tidman} \&
  {Krall}}{1971}]{Tidman1971}
{Tidman} D.~A.,  {Krall} N.~A.,  1971, {Shock waves in collisionless plasmas}.
Wiley

\bibitem[\protect\citeauthoryear{Wheatley, Samtaney  \& Pullin}{Wheatley
  et~al.}{2009}]{Wheatley2009}
Wheatley V.,  Samtaney R.,   Pullin D.~I.,  2009, \mn@doi [Physics of Fluids]
  {10.1063/1.3194303}, 21, 082102

\bibitem[\protect\citeauthoryear{{Yamada}, {Kulsrud}  \& {Ji}}{{Yamada}
  et~al.}{2010}]{Yamada2010}
{Yamada} M.,  {Kulsrud} R.,   {Ji} H.,  2010, \mn@doi [Reviews of Modern
  Physics] {10.1103/RevModPhys.82.603}, \href
  {https://ui.adsabs.harvard.edu/abs/2010RvMP...82..603Y} {82, 603}

\bibitem[\protect\citeauthoryear{{Zenitani} \& {Miyoshi}}{{Zenitani} \&
  {Miyoshi}}{2011}]{Zenitani2011}
{Zenitani} S.,  {Miyoshi} T.,  2011, \mn@doi [Physics of Plasmas]
  {10.1063/1.3554655}, \href
  {https://ui.adsabs.harvard.edu/abs/2011PhPl...18b2105Z} {18, 022105}

\bibitem[\protect\citeauthoryear{{Zenitani} \& {Miyoshi}}{{Zenitani} \&
  {Miyoshi}}{2020}]{Zenitani2020}
{Zenitani} S.,  {Miyoshi} T.,  2020, \mn@doi [\apjl]
  {10.3847/2041-8213/ab8b5d}, \href
  {https://ui.adsabs.harvard.edu/abs/2020ApJ...894L...7Z} {894, L7}

\makeatother
\end{thebibliography}




\appendix



\section{Parallel shocks} \label{app:par}


A parallel shock is where both the velocity and magnetic field vectors are parallel shock front in both the upstream and downstream states, i.e., $\textbf{v}=(v_x,0,0),\textbf{B}=(B_x,0,0)$. Using these, the shock equations reduce to:

\begin{gather}
    \left[ \rho v_x \right]^u _d = 0 \label{eqn:apppar1} \\
    \left[ \rho v_x v_x + P + B^2/2 \right]^u _d = 0 \label{eqn:apppar2} \\
    \left[ \frac{\gamma}{\gamma-1} \frac{P}{\rho} +\frac{v^2}{2} \right]^u _d = 0 \label{eqn:apppar3} \\
    \left[ B_x \right]^u _d = 0
\end{gather}

The compression across the shock can be defined as $\rho ^d/\rho^u = r$. This leads to the relation that $v_x^d / v_x ^u = 1/r$ (from equation \ref{eqn:apppar1}).

Introducing an upstream Mach number $M=v_x^u/V_s^u$, where $V_s^u=\sqrt{\frac{\gamma P^u}{\rho^u}}$ is the upstream sound speed one can determine the upstream velocity as $v_x ^u = M \sqrt{\frac{\gamma P^u}{\rho ^u}}$. One can express the downstream velocity and density in terms off upstream quantities, namely: $\rho ^d = r \rho ^u$ and $v_x ^d = v_x ^u /r$. Equation \ref{eqn:apppar2} becomes:

\begin{gather}
    \rho ^u M^2 \gamma \frac{P^u}{\rho ^u} +P^u = r \rho ^u \frac{1}{r^2} m^2 \gamma \frac{P^u}{\rho ^u} + P^d \\
    \frac{P^d}{P^u} = R =1 + \gamma M^2 (1-\frac{1}{r})
\end{gather}
Finally, $r$ can be determined using Equation \ref{eqn:apppar3} and some algebra as:
\begin{gather}
    r= \frac{M^2 (\gamma +1)}{2 + M^2 (\gamma -1)}
\end{gather}
Therefore, for given upstream properties, the downstream conditions can be easily calculated. In these simulations, the system is normalised such that the upstream sound speed is one by setting $\rho ^u = 1, P^u = 1/\gamma$. In the shock frame, our upstream and downstream states are given in table \ref{tab:parsf}.

\begin{table}
    \centering
    \caption{Analytically determined upstream and downstream states for a the parallel slow-mode shock.}
    \begin{tabular}{c|c|c|c|c|c|c|c|c}
              & $\rho$ & $P$        & $v_x$ & $v_y$ & $v_z$ & $B_x$ & $B_y$ & $B_z$  \\
\hline
   Upstream   &   $1$  & $1/\gamma$ &  $M$  &   0   &   0   & $B_0$ &   0   &   0    \\
   Downstream &   $r$  & $R/\gamma$ &  $M/r$&   0   &   0   & $B_0$ &   0   &   0
    \end{tabular}
    \label{tab:parsf}
\end{table}



\bsp	
\label{lastpage}
\end{document}